\documentclass[11pt,a4paper]{article}
\usepackage{jcapmod}
\usepackage{bm}
\newcommand{\Mp}{M_{\mathrm{P}}}
\newcommand{\M}{N_{\mathrm{f}}}
\newcommand{\ep}{\epsilon}
\newcommand{\be}{\begin{equation}}
\newcommand{\ee}{\end{equation}}
\newcommand{\bea}{\begin{eqnarray}}
\newcommand{\eea}{\end{eqnarray}}
\newcommand{\eref}{\eqref}
\newcommand{\morth}{m_{\perp}}
\renewcommand{\d}{\mathrm{d}}
\newcommand{\e}[1]{\mathrm{e}^{{#1}}}
\newcommand{\vect}[1]{\bm{\mathrm{{#1}}}}

\newcommand{\fNL}{f_{\mathrm{NL}}}

\newcommand{\phase}{\Pi}
\newcommand{\subphase}{\Pi'}

\DeclareMathOperator{\Or}{O}

\newcommand{\etal}{et al.}
\newcommand{\COBE}{\textsc{cobe}}
\newcommand{\CMB}{\textsc{cmb}}

\notoc

\begin{document}

\title{Evolution of $\fNL$ to the adiabatic limit}

\author[a]{Joseph Elliston,}
\author[a,b]{David J. Mulryne,}
\author[c]{David Seery}
\author[a]{and Reza Tavakol}

\affiliation[a]{School of Physics and Astronomy, Queen Mary University of London, \\
Mile End Road, London E1 4NS, U.K.}
\affiliation[b]{Theoretical Physics Group, Imperial College,
London SG1 2AS, U.K.}
\affiliation[c]{Astronomy Centre, University of Sussex,
Falmer, Brighton BN1 9QH, U.K.}

\emailAdd{J.Elliston@qmul.ac.uk}

\emailAdd{D.Mulryne@qmul.ac.uk}

\emailAdd{D.Seery@sussex.ac.uk}

\emailAdd{R.Tavakol@qmul.ac.uk}

\abstract{We study inflationary perturbations in multiple-field models, for which $\zeta$ 
typically evolves until all isocurvature modes decay---the ``adiabatic limit''.
We use numerical methods to explore the sensitivity of the
local-shape bispectrum
%local nonlinear parameter $\fNL$
to the process by which this limit is achieved, finding an appreciable dependence on
model-specific data such as the time at which slow-roll breaks down or the timescale of reheating.
In models with a sum-separable potential where the isocurvature modes decay before the end 
of the slow-roll phase we give an analytic criterion for the asymptotic value of $\fNL$ to be large.
Other examples can be constructed using a waterfall field to terminate inflation while $\fNL$ is 
transiently large, caused by descent from a ridge or convergence into a valley.
We show that these two types of evolution are distinguished by the sign of the bispectrum, 
and give approximate expressions for the peak $\fNL$.}

\maketitle
       
%--------------------------------------------------------
\section{Introduction} \label{sec:introduction}
%--------------------------------------------------------

In single-field inflation with canonical kinetic terms, the curvature perturbation produced
at horizon crossing is conserved with nearly Gaussian statistics \cite{Lyth:2004gb,Maldacena:2002vr}.
Multiple-field models support a richer phenomenology, driven by a 
flow of power from isocurvature modes into the curvature perturbation.
This flow is sourced dynamically, and where only canonical kinetic terms 
are present the dynamics are determined by the potential.
Therefore, there is some hope that we may use one to learn about the other.

How much information could be extracted? A general potential is a complicated landscape,
and it is unlikely that observations will be sufficient to single out a specific shape.
But by piecing together clues from dynamical evolution it may be possible to obtain information 
about the topography of the landscape in our local neighbourhood.
This is a form of potential reconstruction \cite{Copeland:1993jj,Copeland:1993zn,Lidsey:1995np}.

Sensitivity to dynamical effects is helpful when distinguishing observational outcomes.
Unfortunately, it complicates the task of extracting predictions. 
In principle, the statistics of the curvature perturbation should be tracked until
the time of last scattering---%
where the microwave background anisotropy was imprinted---%
and in our present state of ignorance this is an impossible undertaking.
Therefore, to connect the physics of inflation with observation, we must rely on
\emph{conservation}: if the isocurvature modes are exhausted, quenching the
flow of power into the curvature perturbation, it will cease to evolve. It is the statistics
which apply at the onset of conservation which will be inherited by observable quantities.

This point of view was developed soon after multiple-field models entered the 
literature~\cite{GarciaBellido:1995qq}. For practical purposes we require
a characterization of the conditions under which $\zeta$ becomes constant.
In the absence of isocurvature modes, conservation of $\zeta$ was demonstrated
by Rigopoulos \& Shellard~\cite{Rigopoulos:2003ak}, Lyth, Malik \& Sasaki~\cite{Lyth:2004gb}
and later Langlois \& Vernizzi~\cite{Langlois:2005ii,Langlois:2005qp,Langlois:2006iq,
Langlois:2008vk,Langlois:2010vx} using a gradient expansion.
Christopherson \& Malik
extended these results to
models in which the Lagrangian can be an arbitrary Lorentz-invariant
function of the
scalar field and its first derivatives \cite{Christopherson:2008ry}.
More recently, Naruko \& Sasaki and Gao~\cite{Naruko:2011zk,Gao:2011mz}
applied similar arguments
to higher-derivative models
which preserve second-derivative field
equations, where conservation can be
subtle~\cite{Khoury:2008wj,Baumann:2011dt}.
Weinberg developed a different approach \cite{Weinberg:2004kr,Weinberg:2004kf,Weinberg:2008nf,Weinberg:2008si},
adapting the techniques of Goldstone's theorem to show that $\zeta$ would become 
massless on superhorizon scales, admitting a time-independent solution.
Whether this solution is selected is a model-dependent question.

The statistics of $\zeta$ are fossilized in the radiation fluid at last scattering, and 
its two-point correlations have been studied since their presence was confirmed by
{\COBE} \cite{Bennett:1994gg,Bennett:1996ce}.
More recently, sophisticated Cosmic Microwave Background ({\CMB}) experiments 
have raised the possibility that three- and higher $n$-point correlations may be 
detectable \cite{Komatsu:2001rj,Kogo:2006kh}.
These correlations are interesting because, in principle, they are sensitive
probes of physical processes and dynamical conditions in the early universe
\cite{Maldacena:2002vr,Alishahiha:2004eh,ArkaniHamed:2003uz,Lyth:2005fi,Lyth:2005qk}.
But precisely because of these desirable properties, such observables carry an unavoidable risk:
they may be equally sensitive to \emph{subsequent} dynamics, including the process by 
which $\zeta$ becomes conserved. We may learn important physics from studying the
details of this process---%
but it need have little to do with our theories of the very early
universe, and if we wish to use $n$-point correlations to study these theories then we 
should proceed with caution. For this reason it is important to understand which predictions
of early-universe models can be connected reliably to late-time observations.

Our imprecise knowledge of physics above the TeV frontier means it is not possible 
to give a complete answer. In this paper we pursue a more modest objective.
Focusing on three-point correlations---%
with amplitude measured by $\fNL$ \cite{Komatsu:2001rj}---%
and considering models where the flow of power from isocurvature modes is quenched at or near the
end of inflation, we study how the asymptotic value of $\fNL$ depends on the process by which
the isocurvature modes become exhausted. We refer to this exhausted state as the \emph{adiabatic limit}.
Our arguments are phrased in terms of three-point correlations, but many of our conclusions are 
general and apply to arbitrary $n$-point functions including the two-point function.

\paragraph{Classification of models.}
We restrict attention to models where a significant $\fNL$ is generated dynamically by inflation.
This excludes examples such as the curvaton \cite{Lyth:2001nq,Moroi:2001ct} or modulated 
reheating \cite{Dvali:2003em,Zaldarriaga:2003my} which rely on an inflationary seed perturbation
but amplify it by a noninflationary mechanism. If we disallow noncanonical kinetic terms,
Maldacena's result guarantees that the models of interest must include two or more 
dynamically relevant fields \cite{Maldacena:2002vr}.

Under these conditions, field fluctuations are generated at horizon crossing with almost 
Gaussian statistics \cite{Seery:2005gb,Seery:2006vu,Seery:2008ax}. However,
unlike single-field inflation, these perturbations may cause spatially separated regions 
of the universe to experience different expansion histories. The set of phase space 
trajectories associated with an ensemble of such regions is initially a narrowly collimated 
bundle whose `width' is set by quantum scatter. (We give details in \S\ref{sec:theory}.)
The curvature perturbation is a precise measure of the relative expansion between spatial patches.
Therefore its evolution is a consequence of focusing or defocusing of the bundle:
as nearby patches of the universe evolve towards or away from each other in phase space,
they experience varying expansion rates. In the adiabatic limit, the bundle degenerates to a caustic.
Its width shrinks to zero, and all trajectories converge to a single line.

When does convergence occur? 
The answer is model-dependent, but we can recognize broad classes of behaviour.
\begin{itemize}
	\item
	The potential may contain a focusing region, which enables trajectories to converge `naturally.'
	If inflation ends in the vicinity of this region, an adiabatic limit is automatic. Examples
	include Nflation and related models \cite{Dimopoulos:2005ac,Alabidi:2005qi,
	Vernizzi:2006ve,Alabidi:2006hg,Kim:2006ys,Kim:2006te,Kim:2010ud}.
	However, one is always free to build models in which the adiabatic limit is achieved differently, 
	perhaps by a waterfall transition. If the `natural' limit is employed, there are two relevant questions.

	First, is the limit achieved before the end of inflation? If so, it is not necessary to specify
	details of the subsequent phases. Otherwise, we must choose among the various scenarios 
	for reheating and later dynamics, and the predictions of the model may depend on our choice.

	Second, in the case where an adiabatic limit is reached during inflation, does this occur
	before the slow-roll approximation fails? This raises no issues of principle, but may
	influence how one chooses to study the model; for example, numerical methods may be required.
	We will return to this question in \S\ref{sec:models}.

	\item
	Alternatively, there may be no natural means by which trajectories converge.
	In such models there is no alternative: the inflationary model does not make unambiguous 
	predictions by itself, but only when embedded in a larger scenario which determines
	at least the mechanism by which inflation ends and the universe reheats.
\end{itemize}
None of these observations are new, but their application to 
non-Gaussian statistics has yet to be studied in detail.

Accepting the separate universe principle, to be discussed in \S\ref{sec:theory} below,
one obtains explicit expressions for the $n$-point functions of the curvature 
perturbation~\cite{Lyth:2005fi} which automatically respect these conclusions.
Therefore, concrete predictions can be obtained whenever it is possible to calculate
these expressions accurately until the onset of conservation. However, in many cases 
this ideal procedure is impossible or impractical. Working in a special class of models where
the potential is separable, Meyers \& Sivanandam~\cite{Meyers:2010rg,Meyers:2011mm}
argued that the connected $n$-point correlation functions would be damped towards
slow-roll suppressed values. Our analysis is closely related, but we argue that the value 
achieved in the adiabatic limit need not be especially small \cite{Kim:2010ud}. 
Indeed, in some cases, the adiabatic limit is associated with growth towards the 
asymptotic value, rather than decay.

\paragraph{Objectives.}
In this paper we apply these ideas to the primordial bispectrum, and its amplitude $\fNL$ in the squeezed limit.
There are three principal objectives. First, we illustrate that local $\fNL$ can be sensitive to 
details of when and how the adiabatic limit is reached. Using examples, we demonstrate that---%
even in models where convergence occurs naturally---%
$\fNL$ may depend on the details of this process. Second, the calculations necessary to 
obtain a precise estimate of $\fNL$ can be technical, perhaps requiring recourse to
numerical methods. For some models, simple techniques exist which allow  a qualitative
estimate of the evolution and asymptotic value of $\fNL$. We outline these methods and 
apply them to simple examples. Third, we use numerical methods to perform a detailed
study of the evolution of $\fNL$ in a selection of models. By themselves these calculations 
already reveal interesting patterns of behaviour, but also provide guidance regarding the 
asymptotics of models where an adiabatic limit is reached only through the intervention of
post-inflationary dynamics.

\paragraph{Outline.}
The plan of the paper is as follows. In \S\ref{sec:theory} we discuss the separate universe 
approach to perturbation theory in phase space, and the $\delta N$ formalism. 
In \S\ref{sec:largeNG} we discuss mechanisms for generating large evolving local non-Gaussianity, 
and estimates for the maximum value and its dependence on initial conditions.
\S\ref{sec:analytics} includes a brief account of analytic expressions for estimating $\fNL$,
and discusses conditions necessary for this value to be large. \S\ref{sec:models} reports a 
detailed numerical study of $\fNL$ in a selection of models with two or more fields.
We conclude in \S\ref{sec:conclusions}. An Appendix contains details of some analytic calculations.

%--------------------------------------------------------
\section{Phase space description of slow-roll inflation} \label{sec:theory}
%--------------------------------------------------------

Consider inflation driven by multiple canonical scalar fields $\phi_i$ (where $i = 1, 2, \ldots , \M$),
self-interacting through a potential $W(\phi_i)$. Defining $W_{,i} = \partial W / \partial \phi_i$, 
the scalar equations of motion are 
\begin{equation}
	\label{eq:eoms}
	\ddot{\phi}_i + 3H \dot \phi_i +  W_{,i} = 0 .
\end{equation}
Inflation occurs when $\epsilon \equiv - \dot{H}/H^2 < 1$. Eq.~\eqref{eq:eoms}
generates a $2\M$-dimensional phase space $\phase$. In the ``slow-roll'' limit where 
$\epsilon \ll 1$ there is a dynamical attractor, allowing the decaying mode to be discarded
and restricting evolution to an $\M$-dimensional submanifold $\subphase$
on which (for example) the $\dot{\phi}_i$ are determined in terms of the $\phi_i$.
The growing mode on $\subphase$ satisfies $3H \dot{\phi}_i + W_{,i} = 0$.
In the slow-roll limit it is possible to write $\epsilon$ as a sum of independent contributions 
from each field, yielding $\epsilon = \sum_i \epsilon_i + \Or(\epsilon_i^2)$, where the 
$\epsilon_i$ satisfy
\begin{equation}
	\epsilon_i \equiv \frac{\Mp^2}{2} \left( \frac{W_{,i}}{W} \right)^2 .
\end{equation}
In simple models it may happen that the matrix $\eta_{ij} \equiv \Mp^2 W_{,ij} / W$
also has small components $| \eta_{ij} | \ll 1$.

Density fluctuations are generated by the inflationary background and can be measured 
by the curvature perturbation on uniform density spatial hypersurfaces, denoted $\zeta$.
For cosmological purposes its statistical properties are characterized by low-order
correlation functions, of which the first two are
\begin{subequations}
	\begin{align}
		\label{PS}
		\langle
			\zeta_{\vect{k}_1} \zeta_{\vect{k}_2}
		\rangle
		& \equiv
		(2\pi)^3
		\delta(\vect{k}_1 + \vect{k}_2) P(k_1) ,
		\\
		\label{fNL}
		\langle
			\zeta_{\vect{k}_1} \zeta_{\vect{k}_2} \zeta_{\vect{k}_3}
		\rangle
		& \equiv
		(2\pi)^3
		\delta( \vect{k}_1 + \vect{k}_2 + \vect{k}_3 )
		B(k_1, k_2, k_3) .
	\end{align}
\end{subequations}
The amplitude of the three-point function is usually measured in terms of a 
momentum-dependent parameter $\fNL$, satisfying
\begin{equation}
	\frac{6}{5} \fNL(k_1, k_2, k_3) =
		\frac{B(k_1, k_2, k_3)}
			{\sum_{i < j} P(k_i) P(k_j)} .
	\label{fnl}
\end{equation}
where $i$, $j \in \{ 1, 2, 3 \}$. The problem at hand is to calculate $\fNL$.

\paragraph{Bundles of trajectories.}
A comoving scale $k$ is outside the horizon when $k/aH < 1$, where $a$ is the scale factor
and $H \equiv \dot{a}/a$ is the Hubble parameter. An overdot denotes a derivative with 
respect to time. During inflation $H$ is approximately constant, whereas $a$ is growing
rapidly. Therefore $k/aH$ rapidly becomes negligible a few e-folds after horizon crossing.
Smoothing over a comoving scale somewhat larger than $(aH)^{-1}$, widely separated
spatial patches will locally evolve like an unperturbed universe up to small corrections.
This is the separate universe picture \cite{Lyth:1984gv,Bardeen:1983qw, Wands:2000dp}.

An ensemble of smoothed regions picks out a collection of trajectories in phase space.
If the ensemble is drawn from a spacetime region of finite comoving extent $L$
then we can expect this collection to be narrowly collimated provided $L$ is not too large.%
	\footnote{Comoving quantities such as $L$ are not physical, and can not appear 
	in predictions for observable quantities \cite{Senatore:2009cf}. The discussion 
	in this paper is independent of $L$, but as a point of principle $L$ should be 
	removed from physical quantities by supplying a distribution function for the 
	large-scale modes \cite{Bartolo:2007ti,Seery:2009hs,Seery:2010kh}.
	After doing so, all predictions depend only on physical scales. Similar conclusions 
	have been obtained by a number of different methods \cite{Rajaraman:2010zx,Urakawa:2010kr}.}
We refer to this ensemble of clustered trajectories as a bundle. When making predictions 
for microwave background scales it may contain of order $10^6$ trajectories or more \cite{Lyth:2006gd}.
In practice the analysis is simplified by working in a thermodynamic limit where the bundle 
formally contains an infinite number of trajectories. To avoid spurious infrared problems 
we should demand that these reheat almost surely in the same vacuum.

The isocurvature modes label Fermi normal coordinates on $\subphase$, adapted to the bundle.
Each isocurvature field $s$ has equation of motion $\dot{s} = 0$ and constitutes a conserved 
quantity \cite{Salopek:1995vw,GarciaBellido:1995qq}. Together, these conserved quantities 
identify a trajectory. A particular location on each trajectory is specified by the integrated 
expansion $N \sim \ln a(t)$. This `trajectory' approach can be
traced to Hawking's formulation 
of perturbation theory \cite{Hawking:1966qi}, and
was applied to inflation by several authors
\cite{Lyth:1984gv,Starobinsky:1982ee,Starobinsky:1986fxa,Sasaki:1995aw}.
An explicit description in terms of trajectories on $\Pi'$ was given by Salopek \cite{Salopek:1995vw}
and Garc\'{\i}a-Bellido \& Wands \cite{GarciaBellido:1995qq}. Locally, $N$ and the isocurvature fields
generate a coordinate chart on $\subphase$. Fluctuations along the \emph{same} trajectory generate 
the adiabatic mode, $\zeta = \delta N$. Isocurvature modes represent fluctuations \emph{between} 
trajectories. In an $\M$-field model, the reduced phase space supports $\M-1$ isocurvature modes.

\paragraph{Bundle sections.}
Consider a foliation of $\subphase$ by submanifolds which are nowhere tangent to the bundle. 
An important example is foliation by surfaces of fixed energy density. We may use any such 
foliation to replace $N$ as a label for length along the trajectories. Intersecting the bundle 
with an individual hypersurface generates a cross-section with coordinates inherited from 
the isocurvature modes. For example, working on uniform density hypersurfaces, the scalar 
fields take values $\phi_i^c$. These ``coordinates'' are not independent but are subject to the 
constraint $\d W(\phi_i^c) = 0$, leaving the expected $\M-1$ isocurvature labels.

If the bundle has degenerated to a caustic then each hypersurface intersects the bundle at a 
unique point $\phi_i^c$. Making a small change of trajectory $\delta \phi_j^*$ earlier in the 
bundle's evolution, one will observe no change in $\phi_i^c$. Therefore $\delta \phi^c_i = 0$
and we conclude $\partial \phi^c_i / \partial \phi_j^* = 0$. One can regard this as a requirement
that physical predictions computed from statistical properties of the bundle become independent
of its initial conditions. Speaking loosely, we describe this behaviour as an ``attractor.''
The attracting trajectory is the caustic, and we will sometimes refer to it as the \emph{limiting trajectory}.
An ensemble of smoothed patches traversing the limiting trajectory differ only by their relative position
within it, making $\zeta$ conserved.

This argument identifies regions where $\partial \phi^c_i / \partial \phi_j^* \rightarrow 0$
with regions where $\zeta$ is conserved.%
	\footnote{The condition that $\partial \phi^c_i / \partial \phi_j^* \rightarrow 0$ 
	is sufficient to show that that $\zeta$ becomes conserved within the separate universe
	picture, but this argument does not demonstrate that it is necessary.}
We describe it as the \emph{adiabatic limit}. In what follows we will see that estimates 
of the decay rate of $\partial \phi^c_i / \partial \phi^*_j$ play an important role in analysing 
the adiabatic limit. As an example, consider the purely scalar dynamics associated with
slow-roll inflation. A common type of limiting trajectory lies on a valley floor in the landscape 
generated by the potential. Descending into the valley, the mass-squared matrix associated 
with perturbations orthogonal to the direction of motion should be nondegenerate, with all 
eigenvalues large and positive. Taking the smallest eigenvalue of magnitude $\sim \morth$
one will generically find $\partial \phi_i^c / \partial \phi^*_j \sim \e{-\morth^2 N / 3H^2}$.
We will discuss this example more carefully in \S\ref{sec:analytics}.

\paragraph{Bundle averages.}
These principles give a procedure to determine the statistics of $\zeta$ in an adiabatic limit.
One uses the bundle of trajectories to determine any required $n$-point correlation functions, 
and then imposes the requirement $\partial \phi_i^c / \partial \phi_j^* \rightarrow 0$.
Unfortunately there is no unique way to do so. Different approaches to the limit correspond to 
different focusing mechanisms. As we have explained, focusing regions may occur naturally 
in some models; the examples studied by Meyers \& Sivanandam 
\cite{Meyers:2010rg,Meyers:2011mm} are of this type. But whether or not a model naturally 
contains an adiabatic limit we may usually elect to impose a different one, perhaps by
enlarging the field content to include a waterfall transition. We will study some possibilities below.

It is first necessary to obtain the relevant correlation functions. When the slow-roll attractor is 
operative, the e-foldings along each trajectory can be expressed as a function of its initial conditions.
Therefore $N = N(\phi_i^*)$. Expanding in the neighbourhood of a fiducial trajectory yields
\begin{equation}
	\label{eq:deltaN}
	\zeta
	\equiv
	\delta N
	= \sum_{i} N_{,i} \delta \phi_i^*
		+ \frac{1}{2} \sum_{i, j} N_{,ij} 
		\delta \phi_i^* \delta \phi_j^* + \cdots ,
\end{equation}
where $N_{,i} \equiv \partial N / \partial \phi_i^*$ and 
$N_{,ij} \equiv \partial^2 N / \partial \phi_i^* \partial \phi_j^*$.
The $\delta \phi_i^*$ measure deviations from the fiducial trajectory, and will typically be 
of order the quantum scatter. After restriction to connected correlation functions there is 
no dependence on the arbitrary choice of fiducial trajectory.

Eq.~\eqref{eq:deltaN} enables the low-order correlation functions to be expressed in terms 
of the data $N_{,i}$, $N_{,ij}$, which can be computed in some 
models~\cite{Vernizzi:2006ve,Wang:2010si,Meyers:2010rg}. One finds 
$\langle \zeta \zeta \rangle = N_{,i} N_{,j} \langle \delta \phi_i \delta \phi_j \rangle_*$,
which determines the power spectrum~\eqref{PS}. Similarly, $\fNL$ can be written%
	\footnote{We are neglecting intrinsic non-Gaussianities of the $\delta \phi_i^*$.
	For slow-roll inflation with canonical kinetic terms these are negligible whenever
	$\fNL$ is large enough to be observable \cite{Lyth:2005qj,Vernizzi:2006ve}.}
\cite{Lyth:2005fi}
\begin{equation}
	\label{eq:fnl}
	\fNL =
		\frac{5}{6}
		\frac{ \sum_{i,j} N_{,i}  N_{,j}  N_{,ij} }
			{\left( \sum_{i} N_{,i}^2 \right) ^2} .
\end{equation}

%-----------------------------------------------------------
\section{Transitory behaviour of $\fNL$} \label{sec:largeNG}
%-----------------------------------------------------------

It was explained in \S\S\ref{sec:introduction}--\ref{sec:theory} that our interest lies 
in adiabatic regions where all isocurvature modes are exhausted, preventing further 
evolution of $\zeta$. This limit need not be achieved smoothly. For example,
in hybrid scenarios the inflationary phase is suddenly destabilized by a waterfall transition, 
leading to abrupt convergence of trajectories. Although a convincing demonstration 
has not yet been given, in some circumstances the subsequent dynamics may preserve
the value of $\fNL$ at the waterfall. This strategy has been invoked by various authors
\cite{Sasaki:2008uc,Naruko:2008sq,Byrnes:2008wi,Byrnes:2008zy}. Some fine-tuning 
would be required to arrange $|\fNL| \gg 1$ at the transition point. Nevertheless,
in these scenarios and others it may be misleading to focus exclusively on regions where 
phase space trajectories naturally converge. For this reason we pause to study the 
qualitative evolution of $\fNL$, whether or not we are close to an adiabatic region.
We focus on scenarios where its value changes rapidly, before returning to focusing 
regions in \S\ref{sec:analytics}.

Under which circumstances should we expect the moments of the distribution function 
to change significantly? The distribution function describes how trajectories cluster 
around the core of the bundle. It is conserved under linear evolution, but is sheared
or distorted on curved paths \cite{Mulryne:2009kh,Mulryne:2010rp}. These effects 
reshape the distribution function: even when it is initially Gaussian we expect probability 
to be relocated from the core to the outer layers of the bundle. This is associated with 
the generation of significant third- or higher $n^{\mathrm{th}}$-order moments.

Curved paths can be generated by many choices of microphysics. We study only an 
especially simple mechanism. Where the potential's topography includes a ridge 
or valley we will encounter curved trajectories diverging from the ridge or converging 
into the valley floor. Examples of such trajectories have been studied by a number of authors
\cite{Alabidi:2006hg,Byrnes:2008wi,Byrnes:2008zy,Peterson:2010mv}.

%----------------------------
\subsection{Ridges: Diverging trajectories}
\label{sec:ridges}
%----------------------------

We restrict attention to two-field models, which already capture the principal dynamical features,
and label $\subphase$ by coordinates $\phi$ and $\chi$. We assume a ``ridge'' or separatrix 
at $\chi = 0$. In the neighbourhood of an arbitrary point $(\phi_0,0)$ on the ridge the potential 
will generically have the form $W \approx W_0 + g_0 (\phi - \phi_0) - \frac{1}{2}m_\chi^2 \chi^2$.
The mass-squared $m_\chi^2$ is positive, and omitted terms are higher-order in $\phi-\phi_0$ and $\chi$.
These become relevant at some point after the trajectory has been ejected from the vicinity of $\chi = 0$.
The trajectory $\chi = 0$ is classically stable, although depopulated by quantum fluctuations
\cite{Lyth:2010zq,Fonseca:2010nk,Abolhasani:2010kr,Gong:2010zf,Mulryne:2009ci}.

\paragraph{Trajectories.}
Measuring length along each trajectory by the energy density, the evolution equations are
\begin{subequations}
	\begin{align}
		\frac{1}{3\Mp^2}
		\frac{\d \phi}{\d (H^2)}
		& =
			\frac{g_0}{g_0^2 + (m_\chi^2 \chi)^2}
			,
			\label{eq:phi-trajectory}
		\\
		- \frac{1}{3\Mp^2}
		\frac{\d \chi}{\d (H^2)}
		& =
			\frac{m_\chi^2 \chi}
			{g_0^2 + (m_\chi^2 \chi)^2}
			\label{eq:chi-trajectory}
		.
	\end{align}
\end{subequations}
According to~\eqref{eq:chi-trajectory}, a trajectory emanating from $(H_*, \chi_*)$ 
and evolving to $(H_c, \chi_c)$ satisfies
\begin{equation}
	\frac{m_\chi^2}{2} ( \chi^2_c - \chi^2_* ) +
	\frac{g_0^2}{m_\chi^2} \ln \frac{\chi_c}{\chi_*}
	=
	3 \Mp^2 (H^2_* - H^2_c) .
	\label{eq:chi-dispersion}
\end{equation}
If $|\chi_c| \lesssim |g| / m_\chi^2$ then the logarithm dominates and the trajectories 
disperse linearly in the sense $\chi_c = \chi_* D$, where the growth factor $D$ satisfies 
$D \equiv \e{\beta (H_\ast^2 - H_c^2)}$ and $\beta \equiv 3 (\Mp m_\chi / g_0)^2$.
Nonlinear dispersion occurs in the region $|\chi_c| \gtrsim |g_0|/m_\chi^2$ where the 
quadratic term dominates. The transition between the two is the ``turn,'' beyond which 
each trajectory is ejected from the ridge and rapidly evolves to $|\chi_c| \gg |g_0|/m_\chi^2$.
At the turn we have
\begin{equation}
	|\chi_{\text{turn}}| \sim \frac{|g_0|}{m_\chi^2} ,
	\label{eq:chi-turn}
\end{equation}
which makes the kinetic energy in each field roughly equal,
$|\dot{\phi}_{\text{turn}}| \approx |\dot{\chi}_{\text{turn}}|$.

This leads to the following physical picture. Trajectories which are still close to the ridge
preserve their initial Gaussian profile. Trajectories populating the downhill-edge of the bundle
quickly slide away, generating a heavy tail at large $|\chi_c|$. In this region kinetic energy 
has greater relative importance, slowing the expansion rate and enhancing the frequency 
of excursions to large negative $\delta N$. Therefore this mechanism will generate negative $\fNL$
from a Gaussian distribution. 

Whether a large negative amplitude is achieved in practice depends on the initial distribution
of trajectories within the bundle, the nonlinear relation between the fields and $\zeta$,
and for how long the mechanism operates. Sufficiently far down the ridge the trajectories depend
on the completion of $W$.
Therefore, the approach to an adiabatic limit cannot be described 
by the techniques of this section.

\paragraph{$\delta N$ analysis.}
We now translate to $\zeta$ and repeat the analysis in the language of the $\delta N$
method \cite{Lyth:2005fi}. Consider two trajectories originating well before the critical 
turning point, but initially separated by a distance $(\delta \phi_*, \delta \chi_*)$.
It is useful to define $\delta \equiv m_\chi^2 |\chi / g_0|$, where $\delta_* \ll 1$ indicates 
the initial point is very close to the ridge. In this region surfaces of constant energy density
in $\subphase$ practically coincide with surfaces of constant $\phi$. Therefore,
to bring this pair of trajectories to a common energy density $H = H_*$ requires a small 
excess expansion $\delta N \approx (2\epsilon_\phi^*)^{-1/2} \delta \phi_\ast / \Mp $.
The subsequent expansion history, measured to a surface $H = H_c$, can be written 
$N = N(H_c; H_\ast, \chi_\ast)$.

In what follows we work without loss of generality on the positive branch $\chi > 0$, 
and suppose $\phi - \phi_0$ and $\chi$ remain sufficiently small that higher-order terms 
in the potential do not become relevant. Once this assumption fails, $\dot{\phi}$ may 
acquire a nonnegligible dependence on $\chi_*$, potentially invalidating our conclusions.
Passing to the limit where $\delta \phi_*$ and $\delta \chi_*$ become infinitesimal, we conclude
that on arrival at $H = H^c$ the trajectories have experienced expansion histories
which differ by 
\begin{equation}
	\d N \approx
		\frac{3 H_\ast^2}{g_0} \; \d \phi_\ast
		+ 18 m_\chi^4 \Mp^2 \; \d \chi_*
		\int_{H_*^2}^{H_c^2} \frac{H^2 \; \d (H^2)}
			{[ g_0^2 + (m_\chi^2 \chi)^2 ]^2}
		\chi
		\left( \frac{\partial \chi}{\partial \chi_*} \right)_{H_*},
	\label{eq:ridge-delta-n}
\end{equation}
where the partial derivative is to be evaluated at constant $H_*$ and $\chi = \chi(H)$.

Invoking the chain rule, Eq.~\eqref{eq:ridge-delta-n} determines all derivatives of $N$.
We find
\begin{equation}
	N_{,\chi \chi}
	=
	18 m_\chi^4 \Mp^2
	\int_{H_*^2}^{H_c^2} \frac{H^2 \; \d (H^2)}
		{[ g_0^2 + (m_\chi^2 \chi)^2 ]^2}
		\left[
			\frac{g_0^2 - 3(m_\chi^2 \chi)^2}{g_0^2 + (m_\chi^2 \chi)^2}
			\left(
				\frac{\partial \chi}{\partial \chi_*}
			\right)_{H_*}^2
			+
			\chi
			\left(
				\frac{\partial^2 \chi}{\partial \chi^2_*}
			\right)_{H_*}
		\right]
	.
	\label{eq:n-chi-chi}
\end{equation}
So far our considerations have been general. Prior to the turn, Eq.~\eqref{eq:chi-dispersion} makes
$\partial^2 \chi / \partial \chi_*^2$ negligible whereas $\partial \chi / \partial \chi_* \approx \chi / \chi_*$
is exponentially growing. In this region our assumptions make the integrands of
Eqs.~\eqref{eq:ridge-delta-n} and~\eqref{eq:n-chi-chi}
positive, and therefore both $N_{,\chi}$ and $N_{,\chi\chi}$ are negative and decreasing.

If $m_\chi$ is not too small, the integrals of~\eqref{eq:ridge-delta-n} and~\eqref{eq:n-chi-chi}
are dominated by their upper limits---where the exponential growth is maximized. 
Taking the initial evolution in $\chi$ to be almost negligible, this requires
\begin{equation}
	m_\chi^2 \gg \frac{3 \epsilon_* H_*^2}{1-(H_c/H_*)^2}
	\approx 3 \epsilon_* H_*^2 ,
	\label{eq:mass-condition}
\end{equation}
where the approximate equality applies if $H_c$ is at least a little smaller than $H_*$.
If $\chi$ is to be sufficiently light to acquire a quantum fluctuation then $m_\chi \ll H_*$, 
and if both conditions are to be compatible we must require $\epsilon_* \ll 1$. A short calculation yields
\begin{subequations}
\begin{align}
	N_{,\chi} & \approx
	- \frac{3 m_\chi^2 H_c^2}{g_0^2} \chi_*
	\left(
		\frac{\chi_c}{\chi_*}
	\right)^2
	\label{eq:n-chi-estimate}
	\\
	N_{,\chi\chi} & \approx
	\frac{N_{,\chi}}{\chi_*} .
	\label{eq:n-chi-chi-estimate}
\end{align}
\end{subequations}
This relation between $N_{,\chi}$ and $N_{,\chi\chi}$ is a
consequence of the exponential growth of $\chi$ prior to the turn.

Initially, $N_{,\chi}$ and $N_{,\chi\chi}$ are small in comparison with $N_{,\phi}$ 
and $N_{,\phi\phi}$. In addition, $N_{,\phi\chi} \approx \delta_* / \Mp^2$ is constant 
and can safely be neglected. Therefore $\zeta$ is dominated by the fluctuation in $\phi$,
which is practically Gaussian. Using~\eqref{eq:fnl}, we find
\begin{equation}
	\frac{6}{5} \fNL \approx
		\left[
			2 \epsilon_*
			+ \left(
				\frac{N_{,\chi}}{N_{,\phi}}
			\right)^3
			\frac{m_\chi^2}{3 H_*^2} \frac{1}{\delta_*}
			+ \Or\Big(
				\frac{N_{,\chi}}{N_{,\phi}}
				\delta_*
			\Big)
		\right]
		\left[
			1 + \frac{N_{,\chi}^2}{N_{,\phi}^2}
		\right]^{-2} .
	\label{eq:fnl-peak-approach}
\end{equation}
While $|N_{,\chi}| \ll |N_{,\phi}|$, the first term dominates 
and~\eqref{eq:fnl-peak-approach} gives $|\fNL| \approx \epsilon_* < 1$.
As the trajectory moves away from the ridge the $\chi_*$-derivatives become 
increasingly important whereas the $\phi_*$-derivatives are constant. When 
$|N_{,\chi}|$ and $|N_{,\phi}|$ are comparable, $\fNL$ is dominated by the second term
in~\eqref{eq:fnl-peak-approach}. In virtue of~\eqref{eq:mass-condition} and the 
initial condition $\delta_* \ll 1$, this is much larger than $\epsilon_*$ and causes a 
spike in $\fNL$. Estimating the peak to occur when $N_{,\phi} \approx -N_{,\chi}$, we find
\begin{equation}
	\left. \fNL \right|_{\text{peak}}
	\approx
	\frac{\eta_{\chi*}}{\delta_*}
	\approx
	- 0.3 \epsilon_*^{1/2} \frac{\Mp}{|\chi_*|} ,
	\label{eq:fnl-peak}
\end{equation}
where $\eta_\chi \approx \Mp^2 W_{,\chi\chi} / W$ is the standard $\eta$-parameter 
associated with $\chi$. In this expression and similar ones below, including 
Eq.~\eqref{eq:valley-universal-peak}, the numerical prefactor is uncertain by an $\Or(1)$
quantity which depends on the precise balance between $N_{,\phi}$ and $N_{,\chi}$ at the peak.

On approach to the spike, Eq.~\eqref{eq:fnl-peak-approach} predicts that $\fNL$ is 
negative and growing like $(\chi_c/\chi_*)^6$. Subsequently, $\chi$ continues to increase
and $|N_{,\chi}|$ eventually dominates $|N_{,\phi}|$. In this region $\zeta$ is composed 
almost entirely of the $\chi$ fluctuation. The non-Gaussianity becomes practically independent 
of $\delta_*$ and decays like $(\chi_c/\chi_*)^{-2}$. These estimates of the growth rate and 
decay rate are valid before the turn, where $\chi_c$ is growing exponentially
as described below Eq.~\eqref{eq:chi-dispersion}.

Dropping numerical factors of order unity and using~\eqref{eq:chi-turn} to estimate 
$\fNL$ when the fiducial trajectory passes the turn, we find 
\begin{equation}
	\left. \fNL \right|_{\text{turn}}
	\sim - \frac{m_\chi^2}{H_c^2} \approx \left. \eta_\chi \right|_c .
	\label{eq:fnl-turn}
\end{equation}
This is much less than Eq.~\eqref{eq:fnl-peak} and therefore occurs some time after the peak $\fNL$ is achieved.
If Eq.~\eqref{eq:fnl-turn} is not invalidated by higher-order terms in the potential, it
implies that the height of the spike is adjustable independently of $\fNL$ on entry or exit.
Since the peak $\fNL$ occurs \emph{before} most trajectories in the bundle reach the turning
point~\eqref{eq:chi-turn}, our analysis will apply provided these higher-order terms become 
relevant only after the turn.

\paragraph{Scaling relations.}
Eqs.~\eqref{eq:fnl-peak-approach} and~\eqref{eq:fnl-peak} give interesting scaling relations 
for the peak $|\fNL|$, and for its growth and decay near the spike. Eq.~\eqref{eq:fnl-peak} 
suggests that the maximum $|\fNL|$ attained during the spike has a practically universal
power-law scaling for any potential which can be approximated by the coefficients 
$g_0$ and $m_\chi$ up to the turn of the trajectories: for such potentials we should expect
$|\fNL| \propto |\chi_*|^{-\nu}$ with exponent $\nu \approx 1$. In \S\ref{sec:models} we will 
use numerical methods to study models which exhibit this scaling behaviour.

\subsection{Valleys: Converging trajectories}
The converse process occurs when a trajectory approaches a valley, where a bundle of 
trajectories is nonlinearly focused rather than defocused. As above, we specialize to a 
two-dimensional field space labelled by coordinates $(\phi, \chi)$ and suppose there exists 
a valley aligned with the $\chi$ direction. In the neighbourhood of the valley we write
$W \approx W_0 + W_\phi + W_\chi$, where $W_0$ is a constant and
\begin{subequations}
	\begin{align}
		W_\phi & =
			\frac{1}{2} m_\phi^2 \phi^2 \\
		W_\chi & =
			g_0 \chi + \frac{1}{2} m_\chi^2 \chi^2 .
	\end{align}
\end{subequations}
If $m_\phi \gtrsim m_\chi$ then the slopes will be relatively steep in comparison with the 
valley floor. Omitted terms are higher order in $\phi$ and $\chi$, but become increasingly 
irrelevant as $\phi$, $\chi \rightarrow 0$. Sufficiently far from $\phi=0$
the motion is almost entirely in the $\phi$ direction.

During descent into the valley, trajectories populating the uphill edge of the bundle 
experience a larger velocity in the orthogonal $\chi$ direction compared to those lower 
down the slope. Therefore the uphill edge is compressed towards the centroid, generating a  
nonlinear distribution. The tail of the distribution is again on the downhill side, but in this 
case the tail has \emph{lower} kinetic energy and enhances the frequency of excursions to 
a large \emph{positive} $\delta N$. Therefore this mechanism generates a positive $\fNL$.

\paragraph{$\delta N$ analysis.}
The evolution equations are
\begin{subequations}
	\begin{align}
		\frac{1}{3\Mp^2}
		\frac{\d \phi}{\d (H^2)}
		& =
			\frac{W_\phi'}
				{(W_\phi')^2 + (W_\chi')^2}
			,
			\label{eq:phi-trajectoryVal}
		\\
		\frac{1}{3\Mp^2}
		\frac{\d \chi}{\d (H^2)}
		& =
			\frac{W_\chi'}
				{(W_\phi')^2 + (W_\chi')^2}
			\label{eq:chi-trajectoryVal}
		.
	\end{align}
\end{subequations}
In analogy with the ridge case, it is helpful to define a dimensionless measure of distance, 
$\delta$, from the valley floor. We choose $\delta \equiv W_\phi'/W_\chi'$, which measures 
the relative partition of kinetic energy between the fields. Our analysis applies when the 
trajectories begin from an initial position sufficiently high above the valley, 
where $\delta_\ast \gg 1$ and only the $\phi$-field is in motion. In this regime, 
Eqs.~\eqref{eq:phi-trajectoryVal}--\eqref{eq:chi-trajectoryVal} can be integrated to find
\begin{subequations}
	\begin{align}
		\frac{\phi^2_c}{\phi_*^2}
			& =
			1 + \frac{6 \Mp^2}{m_\phi^2} \frac{H^2_c - H^2_*}{\phi_*^2}
			\label{eq:phi-evolution}
		\\
		\chi_c + \frac{g_0}{m_\chi^2}
			& =
			\left( \chi_* + \frac{g_0}{m_\chi^2} \right)
			\left( \frac{\phi_c}{\phi_*} \right)^{m_\chi^2/m_\phi^2}
			\label{eq:chi-evolution}
	\end{align}
\end{subequations}
up to corrections of relative magnitude $1/\delta^2$. 
Eqs.~\eqref{eq:phi-evolution}--\eqref{eq:chi-evolution} cease to be a good approximation
no later than $\delta \sim 1$, when $\dot{\phi} \sim \dot{\chi}$ and the kinetic energy in 
each field is approximately equal. In typical models this occurs at the turn.

For $\delta \gg 1$, surfaces of constant energy density are practically aligned with surfaces 
of constant $\phi$. Adopting the methods of~\S\ref{sec:ridges}, we bring a pair of nearby 
trajectories separated by the displacement $(\delta \phi_*, \delta \chi_*)$ to a common 
value of $H$, and write the number of e-foldings to a subsequent surface of constant
energy density $H_c$ as $N = N(H_c; H_*, \chi_*)$. Passing to the limit of infinitesimal 
$\delta \phi_*$ and $\delta \chi_*$, and using Eqs.~\eqref{eq:phi-evolution}--\eqref{eq:chi-evolution},
we conclude that on arrival at $H = H_c$, the trajectories have experienced expansion histories 
which differ by
\begin{equation}
	\d N \approx
		\frac{3 H_\ast^2}{m_\phi^2 \phi_*} \; \d \phi_\ast
		- 18 \frac{\Mp^2}{m_\phi^4}
		\frac{\phi_*}{\delta_*}
		\, \d \chi_*
		\int_{H_*^2}^{H_c^2} \frac{H^2 \, \d(H^2)}{\phi^4}
		\left\{
			1
			- \mu \left(
				\frac{\phi^2}{\phi_*^2}
			\right)^{\mu}
		\right\} ,
	\label{eq:valley-delta-n}
\end{equation}
where $\phi$ is to be understood as a function of $H$
and we have introduced the mass ratio $\mu \equiv m_\chi^2 / m_\phi^2 < 1$.
Corrections to Eq.~\eqref{eq:valley-delta-n} are suppressed by $1/\delta^2$.

Eq.~\eqref{eq:valley-delta-n} reproduces many features of the ridge analysis. 
The derivative $N_{,\phi}$ is constant, whereas $|N_{,\chi}|$ is initially zero but growing.
Performing the integral, we find
\begin{equation}
	N_{,\chi} = \frac{1}{2\delta_*} \frac{\phi_*}{\Mp^2}
		\Phi\Big( \frac{\phi^2}{\phi_*^2} \Big) ,
	\label{eq:valley-n-chi}
\end{equation}
where the ``growth factor'' $\Phi(x)$ satisfies
\begin{equation}
	\Phi(x) \equiv - \ln x
		+ ( x^{\mu} - 1 )
		+ \frac{W_0}{W_{\phi\ast}} ( x^{-1} - 1 )
		+ \frac{W_0}{W_{\phi\ast}} 
		    \frac{\mu}{\mu - 1} ( x^{\mu-1} - 1 ) .
	\label{eq:Phi}
\end{equation}
We have assumed $W_0$ dominates $W_\ast$, but the generalization to other cases is straightforward.
At $x = 1$ we have $\Phi(1) = 0$. For $x < 1$ the dominant growing term depends on microphysical
details of the model. Under our assumption $W_0 \gg W_{\phi \ast}$ the dominant growth is initially from $x^{-1}$.
Inflation will not end naturally in a model of this type, so some other exit mechanism must be invoked.
We will see examples of this kind in \S\ref{sec:models}. On the other hand, if $W_0/W_{\phi \ast} \lesssim 1$ 
the logarithm will initially dominate. In either case, the asymptotic growth in the limit $x \ll 1$ is from $x^{-1}$.
Therefore, for a typical model $\Phi(x)$ is a complicated function determined by a competition for dominance
between the various terms. However, remarkably, in many cases the behaviour of $\fNL$
is almost independent of these complicated microscopic details.

Differentiating~\eqref{eq:valley-n-chi}, we find
\begin{equation}
	N_{,\chi\chi} =
		\left\{
			\mu \delta_*
			+ \frac{2}{\delta_*}
				\Delta\Big( \frac{\phi^2}{\phi_*^2} \Big)
		\right\}
		\frac{N_{,\chi}}{\phi_*} ,
	\label{eq:valley-n-chichi}
\end{equation}
where $\Delta(x) \equiv (x-1) \, \d \ln \Phi(x) / \d x$. Eq.~\eqref{eq:Phi} implies that $\Delta$ is growing
as $\phi$ decreases to zero. While $|N_{,\chi}|$ is increasing towards $|N_{,\phi}|$ we find $\fNL$ increases,
achieving a maximum value when $|N_{,\chi}| \sim |N_{,\phi}|$. First, suppose the $\Delta$-dependent term 
is subdominant at this time, which implies $\Delta \lesssim \mu \delta_*^2$. We find
\begin{equation}
	\left. \fNL \right|_{\text{peak}}
	\sim
	\eta_{\chi*} \delta_*
	\approx 0.3
	\epsilon_*^{1/2} \frac{m_\chi^2 \Mp}{g + m_\chi^2 |\chi_*|}.
	\label{eq:valley-universal-peak}
\end{equation}
which is independent of the growth rate~\eqref{eq:Phi} and the mass ratio $\mu$.
In this sense, the maximum value~\eqref{eq:valley-universal-peak} is a ``universal'' phenomenon.
When $|N_{,\chi}| > |N_{,\phi}|$ we find that $\fNL$ decays like $\Phi^{-1}$, at least until 
$\Delta \sim \mu \delta_*^2$, when it may stabilize as we will explain below. In analogy with the ridge, 
this sequence of growth and decay gives rise to a spike in $\fNL$. Ultimately $\phi/\phi_*$ will decrease until 
$\delta \sim 1$, and the subsequent behaviour of $\fNL$ must be determined by different methods, such as 
those described in~\S\ref{sec:analytics}. Written in terms of the dimensionless measure $\delta$,
Eq.~\eqref{eq:valley-universal-peak} coincides with~\eqref{eq:fnl-peak} with the identification
$\delta_{\text{valley}} = 1/\delta_{\text{ridge}}$. In this language, the sign of $\fNL$ is inherited from the 
sign of $\eta_\chi$.

Second, consider the ``nonuniversal'' case where the $\Delta$-dependent term dominates~\eqref{eq:valley-n-chichi}.
In this case, $\fNL$ increases until $|N_{,\chi}| \sim |N_{,\phi}|$, achieving a value \emph{larger} 
than~\eqref{eq:valley-universal-peak}. Its precise value is set by the ratio $\Delta / \delta_*$,
and may depend on details of the potential, including the mass ratio. When $|N_{,\chi}| > |N_{,\phi}|$
the time dependence of $\fNL$ is set by $\Delta / \Phi$. Its precise scaling depends on the dominant
term in $\Phi$. In particular, if $\Phi \sim x^{-1}$ then $\Delta / \Phi$ is approximately constant and $\fNL$ 
does not decay. In such cases, $\fNL$ exhibits a plateau and it may no longer make sense to speak of a spike at all.
After the turn is completed, the nonlinear deformation of the bundle will partially relax, leading to decay of 
$|N_{,\chi \chi}|$. The precise details, including the decay rate, are model-dependent. Eventually the fields 
reach equipartition of kinetic energy and this analysis breaks down. 

%----------------------------------------------------------
\section{Asymptotic behaviour of $\fNL$} \label{sec:analytics}
%-----------------------------------------------------------

Whether a large $|\fNL|$ can be generated during an epoch of slow-roll inflation---perhaps from the 
``spike'' mechanisms described above---is irrelevant unless it can be preserved in some adiabatic limit. 
The methods of~\S\ref{sec:largeNG} are insufficient to resolve this question.

The potential may be such that a focusing region is naturally available. If inflation terminates in 
this region then the transitory evolution of $\fNL$ studied in~\S\ref{sec:largeNG} has no necessary 
connection with its final asymptotic value. In certain circumstances, where the focusing region
can be analysed in detail, a relatively simple statement is possible. These are the scenarios 
studied by Meyers \& Sivanandam \cite{Meyers:2010rg,Meyers:2011mm}.
One might have thought that the final $\fNL$ would depend only on the local shape of the potential 
in the focusing region, which will typically be a stable parabolic minimum. If so, the
asymptotic value of $\fNL$ would be universal among all potentials sharing a similarly-shaped minimum.
However, this is not the case. As we will explain, the asymptotic value of $\fNL$ generally depends on 
properties of the potential far from the focusing region.

If multiple focusing regions are available, one must be selected by a combination of dynamics and 
initial conditions. To determine which possibility should be expected by late-time observers who 
map the anisotropy of the {\CMB} requires an understanding of the infrared structure of the
entire inflating volume \cite{Bartolo:2007ti}. This difficult ``measure problem'' remains unsolved.

If a natural focusing region is not available, or is not selected, then one must be imposed and the 
entire analysis becomes significantly more complicated. In this case the transitory evolution studied
in~\S\ref{sec:largeNG} may become relevant. We will have little to say about this possibility,
although we investigate some numerical cases in~\S\ref{sec:models}.

\paragraph{Natural focusing.}
In this section, we study models where inflation ends in a region of the potential 
where the trajectories are naturally focused. Broadly speaking, two possibilities exist.
\begin{itemize}
	\item The asymptotic value of $\fNL$ generated during focusing may be unobservably small, 
	erasing any transiently large non-Gaussianity generated by spikes or other features. This possibility 
	was emphasized by Meyers \& Sivanandam \cite{Meyers:2010rg,Meyers:2011mm}, who worked with 
	a particular class of separable $\M$-field models to be discussed below. However, other possibilities exist.
	\item It may be possible to make the focusing process itself generate a large $\fNL$ by suitable choice 
	of $W$. An example of such a model was given by Kim {\etal}~\cite{Kim:2010ud}. 
	(Indeed, in this model, $\fNL$ grows sharply during approach to the adiabatic limit.)
\end{itemize}
In principle, the behaviour of $\fNL$ in a focusing region could be determined 
from~\eqref{eq:deltaN}--\eqref{eq:fnl} by imposing the limit $\partial \phi^c_i / \partial \phi^*_j \rightarrow 0$.
Unfortunately, it is not known how to compute the ``$\delta N$ coefficients'' 
$N_{,i}$ and $N_{,ij}$ for an arbitrary model. Therefore a systematic discussion of this limit must
apparently await future analytic developments.

Explicit expressions for the $\delta N$ cofficients are known only in very restricted circumstances.
Formulae for quadratic potentials were discussed by Lyth \& Rodr\'{\i}guez \cite{Lyth:2005fi}, 
Lyth \& Alabidi \cite{Alabidi:2005qi} and Alabidi \cite{Alabidi:2006hg}. Later, 
Vernizzi \& Wands \cite{Vernizzi:2006ve} and Battefeld  \& Easther \cite{Battefeld:2006sz} 
gave expressions for an arbitrary sum-separable potential. Taking $W = \sum_i V_i(\phi_i)$, one finds
\begin{equation}
	\label{eq:ss_Ni}
	N_{,i} = \frac{1}{\Mp^2} \left(
		\left. \frac{V_i}{V_i'} \right|_*
		-
		\sum_k \left. \frac{V_k}{V'_k} \right|_c
		\frac{\partial \phi_k^c}{\partial \phi_i^*}
	\right) ,
\end{equation}
where $\partial \phi^c_k / \partial \phi_i^*$ satisfies
\begin{equation}
	\frac{\partial \phi^c_k}{\partial \phi^*_i}
	=
	- \frac{W_c}{W_*}
	\sqrt{\frac{\epsilon_k^c}{\epsilon_i^*}}
	\left(
		\frac{\epsilon_i^c}{\epsilon^c} - \delta_{ik}
	\right) .
	\label{eq:dphic-dphistar}
\end{equation}
A similar expression for a product-separable potential $W = \prod_i V_i(\phi_i)$ 
was obtained by Choi {\etal} \cite{Choi:2007su}. Comparable results for a general class of
sum- and product-type potentials were given by Wang \cite{Wang:2010si} and are summarized 
in the Appendix. It is also possible to take the Hubble rate to be separable
rather than the potential \cite{Battefeld:2009ym,Byrnes:2009qy}.

\paragraph{Focusing in a valley.}
A typical example of a focusing region is a valley of the potential landscape,
perhaps terminating in a local minimum. For $\M$ fields, there 
are at least $\M-1$ heavy directions with masses greater than the Hubble rate.
Quantum fluctuations are suppressed in these directions, which prevent the bundle 
from diffusing up the sides of the valley. The steep slopes cause exponential convergence, 
and rapidly focus the bundle to a line.

In the neighbourhood of the valley floor, we assume it is possible to choose
coordinates on field space for which the potential approximately separates
\begin{equation}
	W \approx V_\varphi(\varphi) + \sum_\alpha V_\alpha(s_\alpha)
	\approx
	V_\varphi(\varphi) + \frac{1}{2} \sum_\alpha m_\alpha^2 s_\alpha^2
	\label{eq:potential-valley}
\end{equation}
where $\varphi$ labels distance along the valley floor---%
which may be a light direction---%
and the $\M-1$ fields $s_\alpha$ are stabilized with masses $m_\alpha = V''_\alpha \gtrsim H$. 
By a suitable choice of coordinates we can arrange that $\langle s_\alpha \rangle = 0$.
To describe a complicated valley it may be necessary to glue several such regions together.
Focusing on the particular region described by~\eqref{eq:potential-valley}, we denote the 
field values on entry to its domain of validity $\bar{\varphi}$ and $\bar{s}_\alpha$.
These will be functions of the initial fields $\phi_i^*$. This initial point could generically occur 
far from the valley, where~\eqref{eq:potential-valley} need not be a good approximation.

The heavy fields $s_\alpha$ evolve according to $3H \dot{s}_\alpha = - m_\alpha^2 s_\alpha$.
After $N$ e-foldings from the point of entry, one finds
\begin{equation}
	s_\alpha = 
		\bar{s}_\alpha(\phi_i^*)
		\e{ - \int_0^N \eta_\alpha(N') \; \d N' } .
	\label{eq:phi-decay}
\end{equation}
The total number of e-folds available within the valley is model-dependent. In a long valley the 
focusing may practically go to completion, making $s_\alpha$ effectively zero. Alternatively, 
if the valley rapidly terminates in a local minimum there may be insufficient time to focus the 
bundle completely.

The fields $\phi_k$ can be written as linear combinations of $\varphi$ and the $s_\alpha$,
giving $\phi_k = \gamma_k \varphi + \sum_\alpha \beta^\alpha_k s_\alpha$.
The $\gamma_k$ and $\beta^\alpha_k$ are constants, which depend only on the choice of 
separable coordinates used in~\eqref{eq:potential-valley}. They are independent of the entry 
point $( \bar{\varphi}, \bar{s}_\alpha )$, which implies
\begin{equation}
	\frac{\partial \phi_k^c}{\partial \phi_j^*} = \sum_\alpha \left( 
	\beta_k^\alpha - \gamma_k \frac{{V_\alpha '}^c}{{V_\varphi '}^c}
	\right) \frac{\partial s_\alpha^c}{\partial \phi_j^*} .
	\label{eq:decay-estimate}
\end{equation}
Therefore $\partial \phi_k^c / \partial \phi^\ast_j$ behaves like a linear combination of derivatives
$\partial s^c_\alpha / \partial \phi_j^\ast$.

The number of e-foldings, $N^c(\phi^*)$, which occur between the entry point 
$( \bar{\varphi}, \bar{s}_\alpha )$ and the surface $c$ will usually depend on the initial point $\phi^*$.
Assuming $N^c(\phi^*)$ does not exhibit a dramatic sensitivity to these initial conditions,
Eq.~\eqref{eq:phi-decay} shows that $\partial s^c_\alpha / \partial \phi_j^\ast$ will
decay exponentially as the trajectory settles into the valley. Potentials may exist which violate this 
condition, but we believe it will be satisfied for a majority of trajectories which flow over reasonably 
smooth potential landscapes. Where it is satisfied, this estimate of the decay rate applies once a trajectory 
has been captured by the focusing region, no matter what form the potential takes globally.

In Eq.~\eqref{eq:decay-estimate} the $\gamma_k$ term will typically decay exponentially, because 
$V_\alpha'^c \sim s_\alpha$ whereas $V_\varphi'^c$ decays less rapidly. Therefore Eq. \eqref{eq:phi-decay} implies
the derivatives $\partial \phi_k^c / \partial \phi_j^*$ decay at least as fast as the lightest isocurvature field.
We conclude%
	\footnote{The asymptotic notation $x \asymp y$	indicates that $x$ and $y$ share a common decay rate.}
\begin{equation}
	\label{eq:deriv_decay}
	\frac{\partial \phi_k^c}{\partial \phi_j^*}
	\asymp
	\e{ - \int_{0}^{N} \eta_s(N') \; \d N' }
	\approx
	\e{ - \eta_s N },
\end{equation}
where $\eta_s = \min \{ \eta_\alpha \}$ and $N$ is the same quantity occuring in Eq.~\eqref{eq:phi-decay}.
The final equality applies if $\eta_s$ is approximately constant during the focusing process.

\paragraph{Separable potentials.}
In a globally sum-separable model, for which $N_{,i}$ satisfies~\eqref{eq:ss_Ni}, it may happen 
that~\eqref{eq:deriv_decay} is sufficiently powerful to make the final ``$c$-term'' irrelevant.
In these circumstances the correlation functions of $\zeta$, including the spectrum and bispectrum, 
can be determined from the remaining term of~\eqref{eq:ss_Ni}, which depends only on boundary 
data at the initial time. For correlation functions among fields carrying comparable momenta of
order $k$ this is often taken to be the horizon-crossing time $|\eta| \sim 1/k$, where $\eta$ is the conformal time.
For this reason, the scheme has sometimes been called the \emph{horizon-crossing approximation}
\cite{Kim:2006ys,Kim:2006te,Kim:2010ud}. Despite the name, we caution that this approximation does 
not consist of assuming that the perturbations are constant after horizon-crossing, but rather that their 
values in the adiabatic limit can be determined in terms of the shape of the potential there. 
A similar procedure can be applied in product-separable cases.

It is less straightforward to estimate the minimum number of e-folds required to make the $c$-terms 
of~\eqref{eq:ss_Ni} negligible. Although Eq.~\eqref{eq:deriv_decay} gives information concerning the
decay rate, the number of e-folds required to damp any contribution from the $c$-terms depends on 
their amplitude on entry to the valley. This is a function of each species' relative contribution to the 
energy density of the universe on the initial and final slices $c$ and $\ast$, from which it does not 
appear straightforward to draw general conclusions. However, since the isocurvature masses should be
comfortably heavier than the Hubble scale, the parameter $\eta_s$ will typically be much larger than
unity. In these circumstances, rather less than $\Or(10)$ e-foldings are usually required to accumulate a very 
substantial suppression of the $c$-terms.

In the language of Meyers \& Sivanandam \cite{Meyers:2010rg,Meyers:2011mm}, this damping of the 
$c$-terms is precisely the exponential suppression which they suggested would drive the bi- and 
trispectrum to slow-roll suppressed values. In the language of \S\ref{sec:theory} it represents focusing of the
bundle to a caustic. Our analyses are entirely consistent, but it is helpful to recall that the $\ast$-term 
in~\eqref{eq:ss_Ni} need not be especially small. We briefly comment on this possibility at the end of this section.
If that is the case, suppression of the $c$-terms can cause the correlation functions to experience a short phase
of exponential growth as they approach their asymptotic values.
Note that
all these conclusions depend on the existence of a globally separable potential. We are not aware of a 
systematic study of the asymptotics of $\fNL$ in more general cases.

One might harbour some reservations that the $c$-terms do not decay if the fields settle into a stable 
minimum, for which $V_k / V_k'$ diverges. Near an arbitrary point, which can be chosen as the origin 
without loss of generality, $V_k$ can be written $V_k \approx A + B \phi_k$ and $V_k/V_k'$ approaches a constant.
Near a minimum, one finds instead $V_k \approx A + B \phi_i^2$. Therefore
\begin{equation}
	\frac{V_k}{V_k'}
	\frac{\partial \phi_k^c}{\partial \phi_j^*}
	\approx
	\frac
	{A + B
		\left(
			\gamma_k \varphi^c + \sum_\alpha \beta^\alpha_k s_\alpha^c
		\right)^2
	}
	{2 B
		\left(
			\gamma_k \varphi^c + \sum_\alpha \beta^\alpha_k s_\alpha^c
		\right)
	}
	\sum_\rho
	\left( 
		\beta_k^\rho - \gamma_k \frac{{V_\rho '}^c}{{V_\varphi '}^c}
	\right)
	\frac{\partial s_\rho^c}{\partial \phi_j^*} .
	\label{eq:minimum-decay-estimate}
\end{equation}
If $A = 0$ the prefactor decays. Since the potential is sum-separable, we may always redefine all 
but one $V_k$ to satisfy this condition. However, if the potential is not zero at the minimum then the
remaining $V_k$ must have nonzero $A$. Eq.~\eqref{eq:minimum-decay-estimate} shows that we 
should choose the field $\phi_k$ to have nonzero overlap with the direction of the valley floor, ie., $\gamma_k \neq 0$.
Under these circumstances the right-hand side of~\eqref{eq:minimum-decay-estimate} still decays 
(although perhaps at a reduced rate), because by assumption $\varphi$ decays strictly more slowly than any
isocurvature mode.

\paragraph{Large non-Gaussianity after natural focusing.}
Is it possible to obtain large $\fNL$ at the adiabatic limit?
Working in a sum-separable potential, the foregoing discussion implies that the $c$-dependent terms 
in~\eqref{eq:ss_Ni} may be discarded provided enough focusing can be achieved before the end of inflation.
In general, the conditions required to achieve large $\fNL$ may still be complicated. However, a relatively 
simple picture emerges if we assume that $N_{,i}$ is large for one field $\phi$ (or at most a few such fields) 
\cite{Kim:2010ud}. Therefore, $V_\phi/V_\phi'$ at horizon crossing dominates the analogous terms for all other 
fields and $\fNL$ can be written 
\begin{equation}
	\fNL \approx - \frac{5}{6} \Mp^2
		\left. \frac{V_\phi''}{V_\phi} \right|_* .
	\label{eq:hilltop-fnl}
\end{equation}
In a single-field model the quantity $V_\phi''/V_\phi$ would be the inflationary $\eta$-parameter.
But in an assisted inflation the total potential may be much larger than $V_\phi$ \cite{Liddle:1998jc,Green:1999vv}.
Therefore $\eta_\phi$ can remain small, making $\phi$ light at horizon crossing and causing it to acquire a 
quantum fluctuation by the usual mechanism, while $V_\phi''/V_\phi$ can be appreciable.
We study an example of this type in \S\ref{sec:large-natural}. In such models the sign of $\fNL$ is inherited
from an ``enhanced'' $\eta$-parameter, as in Eqs.~\eqref{eq:fnl-peak} and~\eqref{eq:valley-universal-peak}, but
unlike these cases the enhancement is measured by the
initial share of the energy density contributed by $\phi$,
rather than the
parameters $\delta_{\mathrm{ridge}}$, $\delta_{\mathrm{valley}}$.
In Eq.~\eqref{eq:hilltop-fnl} this enhancement factor is $W/V_\phi \gg 1$.

If several fields have comparable $N_{,i}$, their perturbations contribute equally to $\zeta$ at the 
adiabatic limit and dilute any non-Gaussianity by the same interference effect which leads to the central limit
theorem. Therefore the largest values of $\fNL$ will be achieved where a single field has a dominant $N_{,i}$.

A similar discussion can be given for product-separable potentials. In this case the formulas depend solely 
on one field, labeled $\phi_k$, which must be the field still evolving at the adiabatic limit. Therefore
$\fNL \approx 2 \ep_k^* - \eta_{k k}^*$, and  a large $\fNL$ would require a violation of slow-roll. 
We conclude that large $|\fNL|$ is not possible at the natural adiabatic limit in this class of models.

%------------------------------------
\section{Models} \label{sec:models}
%-----------------------------------

The results of~\S\ref{sec:analytics} show that, even for models where an adiabatic limit can be 
approached analytically, numerical calculations may be necessary to determine the degree of focusing 
which occurs near the end of inflation. In other cases there is simply no alternative.

In this section, we report the results of numerical simulations and compare the outcome to the analytic 
theory developed in \S\S\ref{sec:largeNG}--\ref{sec:analytics}. In appropriate circumstances we show 
that the simplified description of ``spikes'' obtained in \S\ref{sec:largeNG} is an accurate match for full 
numerical simulations. We give examples where the focusing described in \S\ref{sec:analytics} goes to 
completion---making the ``horizon crossing approximation'' highly accurate---and others where it does
not. A case of special interest occurs when the bundle would focus only slightly after the end of inflation. 
One might expect that the error in analytic predictions based on the strict adiabatic limit would be small, 
but it transpires that $\fNL$ can be rather sensitive to the details of the model. In models where there is 
no natural adiabatic region, reheating must occur before an adiabatic limit is reached. In these cases
we perform a qualitative study of the dependence of $\fNL$ on the details of the reheating phase.

\paragraph{Evolution after slow-roll using $\delta N$.}
The phase space description of inflationary trajectories was discussed in \S\ref{sec:theory}.
During multiple-field slow-roll inflation, each trajectory lies on a submanifold $\subphase$ of the full phase space.

In a model more general than multiple-field slow-roll inflation, extra coordinates will typically be required.
First, if the slow-roll approximation fails then one must work on the full phase space $\phase$ rather than the
attractive submanifold $\subphase$. Therefore new isocurvature modes are typically required to label the 
conjugate momenta $\pi_i \sim \dot{\phi}_i$. Second, matter species other than scalar fields may be included, 
perhaps to describe a phase of reheating. In such cases, the full phase space splits into a Cartesian product
constructed from the phase space for each species, and suitable isocurvature modes labelling all these 
coordinates will be required. For thermalized radiation, a common choice is the temperature, $T$.

Several numerical approaches exist to compute the statistics of the density fluctuation 
\cite{Mulryne:2009kh,Mulryne:2010rp,Peterson:2010mv}. Here we take the simple approach of calculating 
the derivatives of $N$ using a finite difference scheme. This requires the slow-roll approximation at 
horizon crossing, where initial conditions are set, but not subsequently. We have verified that our results 
are insensitive to changes in the step size of the finite difference scheme. Although slower than other 
approaches \cite{Mulryne:2009kh,Mulryne:2010rp}, direct $\delta N$ has the advantage of straightforward
comparison with our analytic methods. Moreover, it requires only the evolution  of an unperturbed universe,
making a simple description of reheating---assuming thermal equilibrium and a single radiation fluid---%
easy to implement. These assumptions are at best quasi-realistic, but serve to indicate a plausible phenomenology.

%--------------------------------------------------
\subsection{Two-field models}
%--------------------------------------------------
\subsubsection{Transitory models with interruption}
%--------------------------------------------------
All two-field models documented in the literature which produce large, transient $\fNL$ exploit the spikes 
described in \S\ref{sec:largeNG} \cite{Alabidi:2006hg,Byrnes:2008wi,Byrnes:2008zy}. 
Eqs.~\eqref{eq:fnl-peak-approach} and~\eqref{eq:fnl-peak} show that to obtain large $|\fNL|$ from a ridge, one must 
tune the initial conditions so that $\chi_* / \Mp \ll 1$. Also, if this large $\fNL$ is to be preserved in the 
adiabatic limit, Eq.~\eqref{eq:fnl-turn} implies that some mechanism must operate to end inflation before
the majority of trajectories in the bundle encounter the turn. In two-field models with separable potentials,
the parameter combinations required to ensure these conditions were given by Byrnes {\etal} \cite{Byrnes:2008wi}.
The observables predicted in such models depend strongly on the choice of exit mechanism. In certain cases, such 
as two-field hybrid inflation, it is possible that $\fNL$ is not erased. In other cases this outcome appears unlikely.

\paragraph{Two-field hybrid inflation.}
This model was studied by Alabidi \& Lyth \cite{Alabidi:2006hg} and later by Byrnes 
{\etal} \cite{Byrnes:2008wi,Byrnes:2008zy}. The potential is
\begin{equation}
	\label{eq:hybrid}
	W
	=
	\frac{1}{2} m_\phi^2 \phi^2
	+ \frac{1}{2} m_\chi^2 \chi^2
	+ \frac{1}{2} \left(
		g_\phi^2 \phi^2 \sigma^2
		+ g_\chi^2 \chi^2 \sigma^2
	\right)
	+ \frac{1}{4} \lambda \left ( \sigma^2 - v^2 \right)^2
\end{equation}
where $\phi$ and $\chi$ are slowly-rolling fields, and $\sigma$ is a waterfall field which becomes 
destabilized when ${g_\phi}^2 \phi^2 + {g_\chi}^2 \chi^2 = \lambda v^2$. We take the masses 
$m_\phi$ and $m_\chi$ to be positive, and assume $g^2_\phi/g^2_\chi = m^2_\phi/m^2_\chi$. This ensures
that the waterfall occurs at fixed energy density, making it unnecessary to account for the effect of 
inflation ending on different hypersurfaces \cite{Alabidi:2006wa,Sasaki:2008uc,Naruko:2008sq,Huang:2009vk}.
If the masses are not equal, there is a steep slope in the direction of the more massive field.
The trajectories evolve along this steep direction and then turn towards the global minimum. We expect
some non-Gaussianity to be generated during this process.

\begin{figure}[t] 
	\center{\includegraphics[width = 8cm]{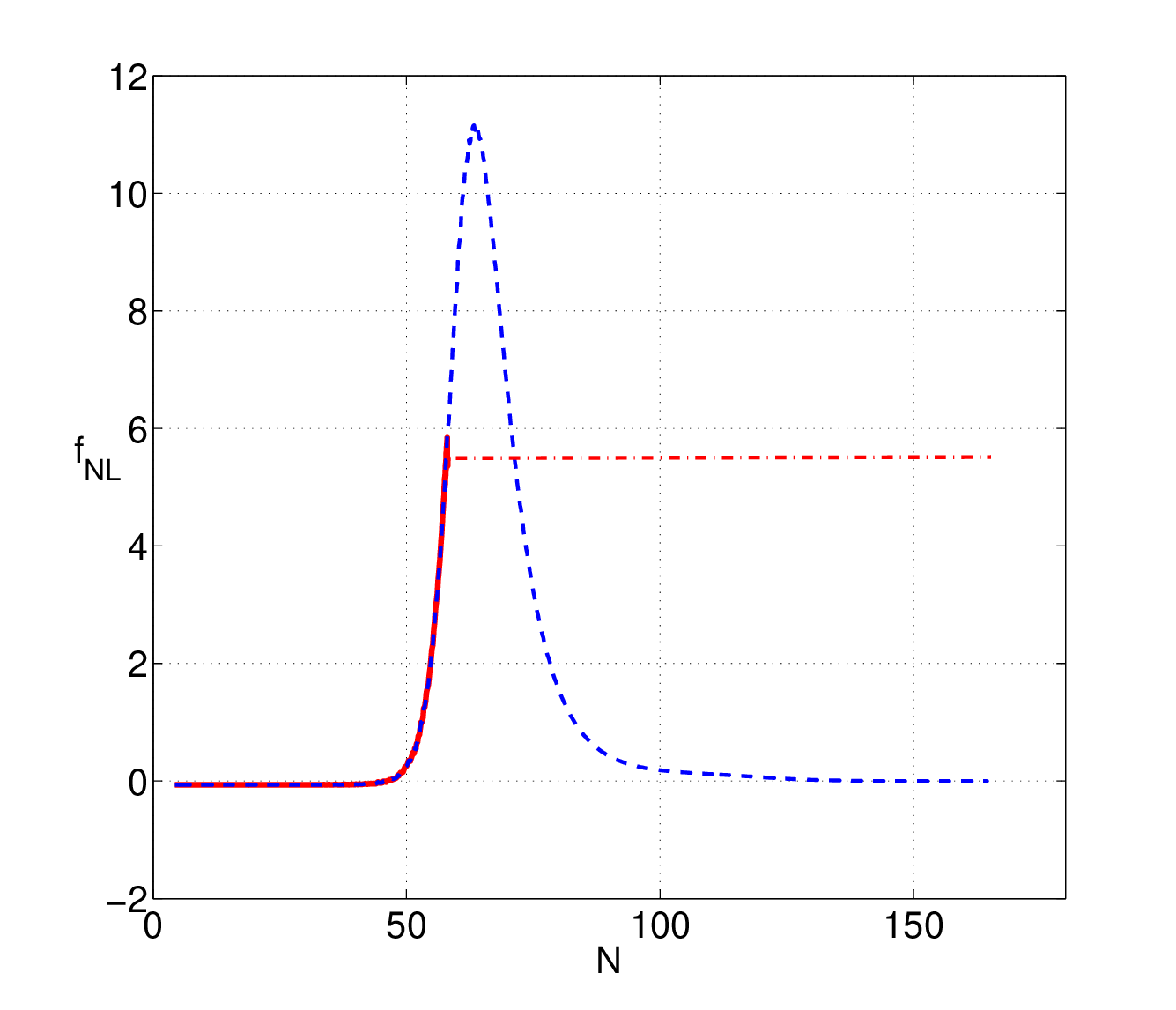}}
    	\caption{Evolution of $\fNL$ for the model \eqref{eq:hybrid} calculated numerically (solid red line)
    	with the hybrid transition included for the parameter values in the text, the (red) dot-dashed line is 
	added to illustrate the nearly constant final level of $\fNL$. The (blue) dashed line, represents the 
	analytical evolution, with no hybrid transition included.}
	\label{figHy}
       \end{figure}

In Fig.~\ref{figHy} we plot the evolution of $\fNL$ for 
$m_\phi/m_\chi = 5$, $\eta_\phi = 4 \Mp^2 m_\phi^2/ (\lambda v^4) = 0.08$ 
and initial conditions $\chi_* = 0.001 \Mp$, $\phi_*=0.5 \Mp$.
We adjust the remaining parameters so that the waterfall occurs when $|\fNL| > 1$ and takes much 
less than a Hubble time to complete. The blue dotted line represents the $\fNL$ generated by the 
slow-roll fields, ignoring the waterfall. If inflation fails to terminate on the spike this leads to a negligible 
asymptotic bispectrum. The potential is of `valley' type with $g=0$, for which the analysis of \S\ref{sec:largeNG} 
explains the positive spike. Beginning from $\fNL \sim \epsilon_*$ there is rapid growth to a positive peak, 
followed by a softer decay. The peak value is well approximated by Eq. \eqref{eq:valley-universal-peak}
which yields $\fNL \approx 8$. By varying the initial conditions, we have confirmed that the peak value 
scales approximately as $1/\chi*$, as predicted by~\eref{eq:valley-universal-peak}.

The solid red line represents a numerical evolution, terminated by a waterfall transition on the growing 
arm of the spike. In the early stages, the numerical results follow the analytic prediction. We continue 
the calculation past the end of inflation by allowing the waterfall field $\sigma$ to become operative.
Care must be taken in modelling this transition. We give the waterfall field a small value consistent with 
the typical  RMS value expected from quantum mechanical excitations of a massive field in de Sitter, 
$\sigma_{\text{RMS}} \approx H^3/M $, where $M$ is representative of the waterfall mass before the transition. 
The results are extremely insensitive to its precise value.%
	\footnote{In reality, this RMS value is made up of many inhomogeneous short scale modes.
	Their collective evolution approximates that of a homogeneous mode, at least in the initial 
	stages before the minimum is reached \cite{GarciaBellido:1996qt}. We expect 
	this approximation captures at least some of the physics which occurs at the hybrid transition.
	}
	
The hybrid field is heavy at horizon crossing and is therefore unperturbed. Hence, we need not differentiate
$N$ with respect to $\sigma$. Using these assumptions, we find that $\fNL$ appears to be conserved through 
the hybrid transition (see Fig.~\ref{figHy}). The numerical evolution is only continued for a fraction of an 
e-fold after the transition, during which time $\fNL$ \emph{does} evolve due to oscillation of the primary fields. However, the
resulting oscillations in $\fNL$ are decaying and are centred around a fixed value. This behaviour is apparently 
generic for a range of parameter values, provided the transition happens sufficiently rapidly---%
in less than an e-fold. One must already impose this ``rapid transition'' condition to avoid issues with primordial 
black holes \cite{GarciaBellido:1996qt}.%
	\footnote{Note that when the hybrid transition does not occur on a uniform density hypersurface 
	a significant extra contribution to $\fNL$ can be generated \cite{Alabidi:2006wa,Huang:2009vk}. We have checked a small number 
	of these cases numerically and find agreement with the formulae given in 
	Ref. \cite{Byrnes:2008zy}, although we have only considered examples of positive 
	curvature rather than ridges.}

%--------------------------------------------------
\subsubsection{Large non-Gaussianity at the natural adiabatic limit}
\label{sec:large-natural}
%--------------------------------------------------

We illustrate this case using a model closely related to the $N$-axion model of Kim {\etal} \cite{Kim:2010ud},
in which the potential is taken to be $V=\sum_i \Lambda_i^4 ( 1-\cos 2\pi f_i^{-1} \phi_i )$. The sum is taken 
over a large number of uncoupled axions, and $f_i$ is the decay constant for the $i^{\text{th}}$ axion.
We will study further examples of this type in \S\ref{sec:n-field-models}. In Ref.~\cite{Kim:2010ud}, many 
axions were invoked to generate a phase of assisted inflation. Because the potential is sum-separable, the 
perturbations can be calculated at the adiabatic limit using~\eqref{eq:ss_Ni} after dropping the $c$-term
\emph{provided} inflation ends when the final field gracefully exits from slow-roll. Whether this occurs
depends on the number of fields and the choice of $f_i$. Taking $f_i = f < \Mp$ for all $i$ and supposing that
the initial conditions are chosen so that only a small number of fields populate the hilltop region near $\phi_i = 0$,
the asymptotic $\fNL$ can be calculated using~\eqref{eq:hilltop-fnl}. It will typically be moderate or large.

In this section we study a related two-field model. Dynamically, the large number of axions which begin away 
from the hilltop region serve only to source the Hubble rate. The single field closest to the hilltop sources 
the non-Gaussianity. (This model has some similarity to the scenario of Boubekeur \& Lyth \cite{Boubekeur:2005fj}.)
Therefore, most of the axions can be replaced by a single effective field with a quadratic potential, retaining the 
full cosine only for the axion closest to the hilltop, 
\begin{equation}
	\label{eq:2-field-axion}
	V= \frac{1}{2} m^2 \phi^2 + \Lambda^4 \left (
		1-\cos \frac{2 \pi \chi}{f}
	\right) ,
\end{equation}
where $\Lambda$ and $f$ are constants.

Near the hilltop, the axion potential approximately satisfies $V(\chi)=2 \Lambda^4 (1 - \pi^2 \chi^2/f^2)$.
This yields a tachyonic mass $2 \pi\Lambda^2/f$. Adjusting the $\phi$ potential if necessary to ensure that 
$\chi$ remains light at horizon crossing, the mass induces a large $\fNL$ via~\eqref{eq:hilltop-fnl}.

\begin{figure}[t] 
	\center{\includegraphics[width = 8cm]{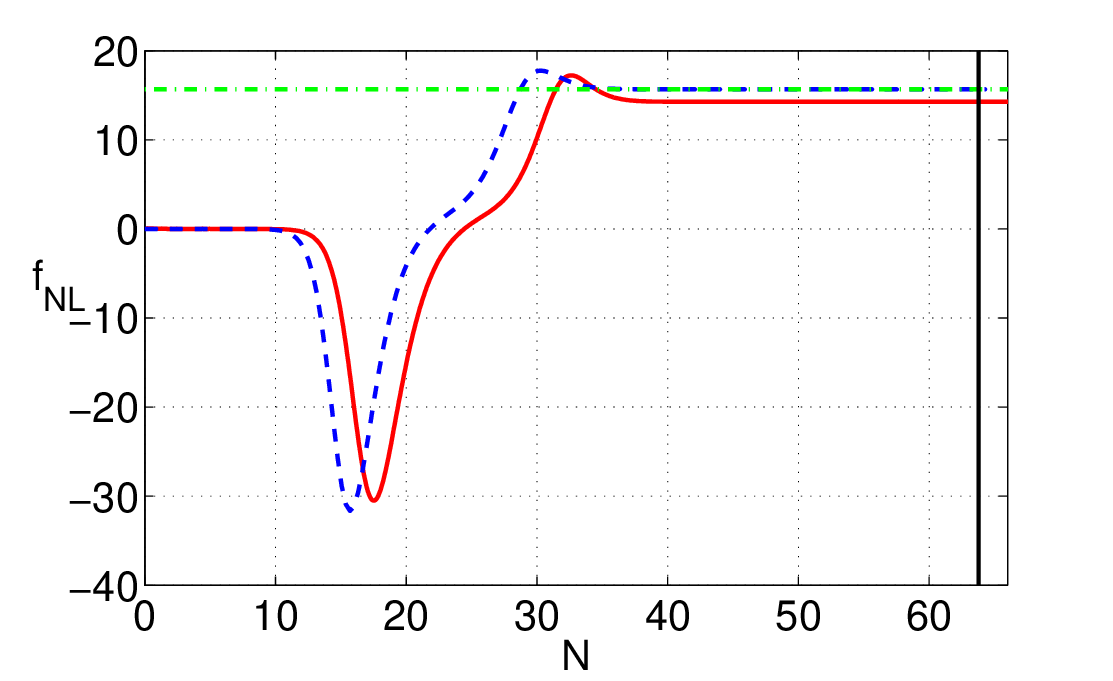}}
	\caption{Evolution of $\fNL$ for the model of Eq.~\eqref{eq:2-field-axion} (first set of parameter choices).
	The solid red line is a numerical calculation. The blue dashed line is an analytic prediction.
	The horizontal green dashed line represents the analytically calculated adiabatic limiting value.}
	\label{figAx1}        
\end{figure}

\begin{figure}[b] 
	\center{\includegraphics[width = 8cm]{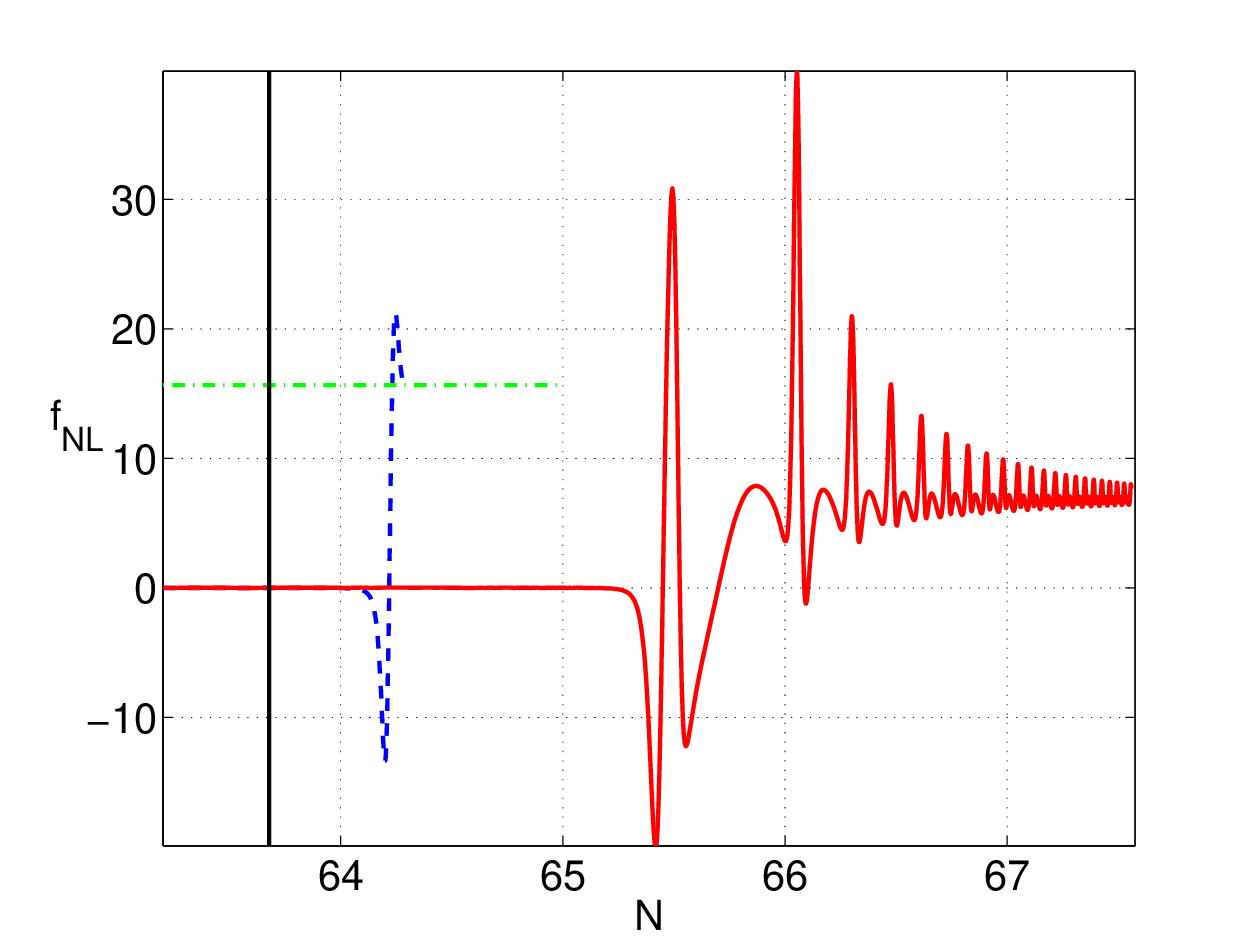}}
	\caption{Evolution of $\fNL$ for Eq.~\eqref{eq:2-field-axion} (second set of parameter choices).
    	The solid red line is a numerical calculation. The blue dashed line is an analytic prediction.
	Only the final few e-folds are shown. The horizontal green dashed line represents the analytically 
	calculated adiabatic limiting value. The solid vertical line indicates when inflation ends,
	computed using the exact equations of motion. As the axion rolls, inflation momentarily restarts 
	and the slow-roll expressions cease to be a good approximation.}
\label{figAx2}        
        \end{figure}

In Figs.~\ref{figAx1} and~\ref{figAx2} we show a numerical evolution for two choices of parameters.
In Fig.~\ref{figAx1} we take $f= \Mp$ and $\Lambda^4 = 25 m^2 f^2/(4\pi^2)$, which makes
the mass of the axion five times greater than the mass of $\phi$. The initial conditions are 
$\phi_*= 16\Mp$ and $\chi_*=(f/2-0.001) \Mp$. As a consequence of its large mass, the axion rolls off 
the ridge quite early. Therefore the system evolves to the limiting trajectory long before the
end of inflation. According to \S\ref{sec:largeNG}, departure from the ridge should produce a large negative spike.
Later, convergence into the minimum should produce a positive spike,
perhaps followed by a plateau. Finally, as the isocurvature modes are exhausted, 
the system should evolve to the adiabatic limit~\eqref{eq:hilltop-fnl}. These features are clearly visible in Fig.~\ref{figAx1}.
Matching the potential~\eqref{eq:2-field-axion} to the analysis of \S\ref{sec:largeNG} and using Eq.~\eqref{eq:fnl-peak}, 
we expect the negative peak of $\fNL$ to occur at $\fNL  \approx - 0.3 \Mp \epsilon_*^{1/2} (f/2 - \chi_*)^{-1} \approx -26$.
This gives good agreement with the observed value. By varying the intial conditions we have verified that scaling with 
$1/(f/2 - \chi*)^{-1}$ is  reproduced to a good approximation. In this case the difference between  our slow-roll analysis 
and the full numerical calculation is at the level of a few percent, consistent with the accuracy of the slow-roll approximation.

%--------------------------------------------------------
%---------------------------------------------------------------
%---------------------------------------------------------

%--------------------------------------------------
In the second example we take $f= \Mp$ and $\Lambda^4 =  m^2 f^2/(4 \pi^2)$, giving both fields the 
same mass. In this case the axion starts to evolve only near  the end of inflation, where $\phi$ is approaching the minimum.
Indeed, much of its evolution takes place while $\phi$ is oscillating. In these circumstances the adiabatic limit cannot be calculated 
analytically using \eqref{eq:ss_Ni}. Nevertheless, the important features can still be understood. While $\phi$ is oscillating 
in the minimum, its potential energy contributes to the energy density in a way not accounted for by the slow-roll approximation.
If we suppose the $\phi$ oscillations do not lead to rapid reheating or preheating, we may expect $\zeta$ to approach a constant
as the trajectories settle in the minimim and Hubble friction drains their energy. The results are given in Fig.~\ref{figAx2}.
In this simple example, $\fNL$ oscillates around an asymptotic value which is lower than would be expected if the adiabatic limit 
were reached during inflation. In more sophisticated examples, where complex dynamical behaviour can occur during the 
oscillating phase, it would be necessary to follow their decay in precise detail \cite{Traschen:1990sw,Kofman:1994rk,Shtanov:1994ce,
Chambers:2007se,Chambers:2008gu,Bond:2009xx,Kohri:2009ac}.

This example is representative of a class of model where natural focusing occurs---%
in this case, caused simply by Hubble damping---%
but does so only after the slow-roll assumption is violated. There are other models in this class which lead to a 
large non-Gaussianity at the natural adiabatic limit, such as models which possess an inflection point in their 
potential with a very slight gradient. We intend to return to these cases in future work \cite{proceedings}.

\subsubsection{Models with no reconvergence in field space}
%--------------------------------------------------

The third possibility discussed in \S\ref{sec:introduction} occurs when the trajectories disperse in field-space 
but the potential provides no region which would enable them to refocus. An example is provided by the model
\begin{equation}
	\label{eq:no-reconverge}
	V= V_0 \phi^2 e^{-\lambda \chi^2} ,
\end{equation}
which was introduced by Byrnes {\etal} \cite{Byrnes:2008wi}. The dynamics were followed only until the 
end of slow-roll, at which time the non-Gaussianity was indeed large. However, at this point, the isocurvature 
modes were not exhausted and the curvature perturbation was still evolving. To study this model, we make the
the same parameter choices as Byrnes {\etal}, 
setting $\lambda=0.05/\Mp^2$, $\phi_i=16\Mp$ and $\chi_i=0.001\Mp$ \cite{Byrnes:2008wi}.

\begin{figure}[b] 
	\center{\includegraphics[width = 8cm]{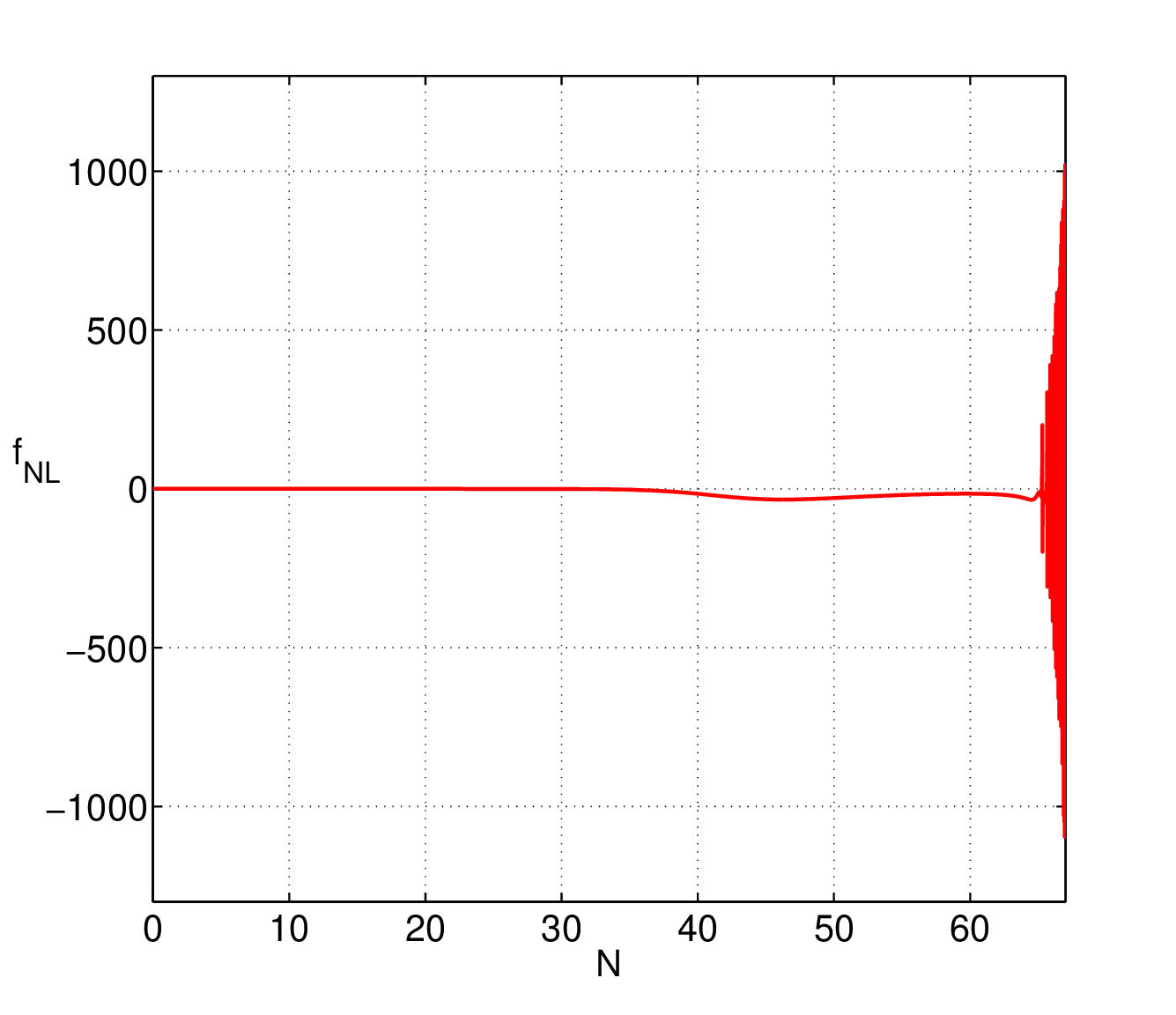}}
    	\caption{Evolution of $\fNL$ for the model \eqref{eq:no-reconverge}.
    	The solid red line is a numerical calculation for the parameter values quoted in the text and $\Gamma = 0$.}
	\label{figChris2}  
\end{figure}

\begin{figure}[t] 
	\center{\includegraphics[width = 8cm]{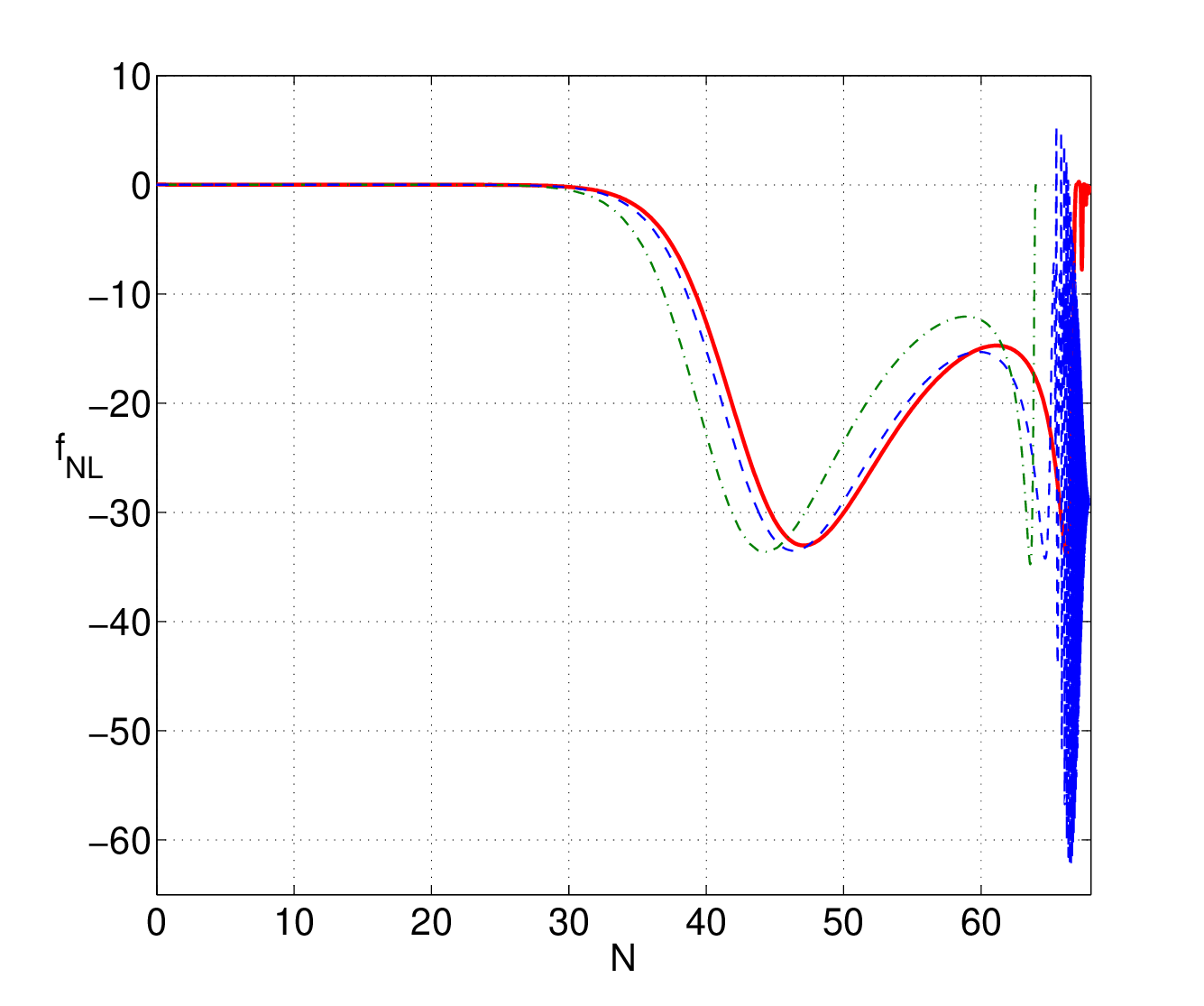}}
 	\center{\includegraphics[width = 8cm]{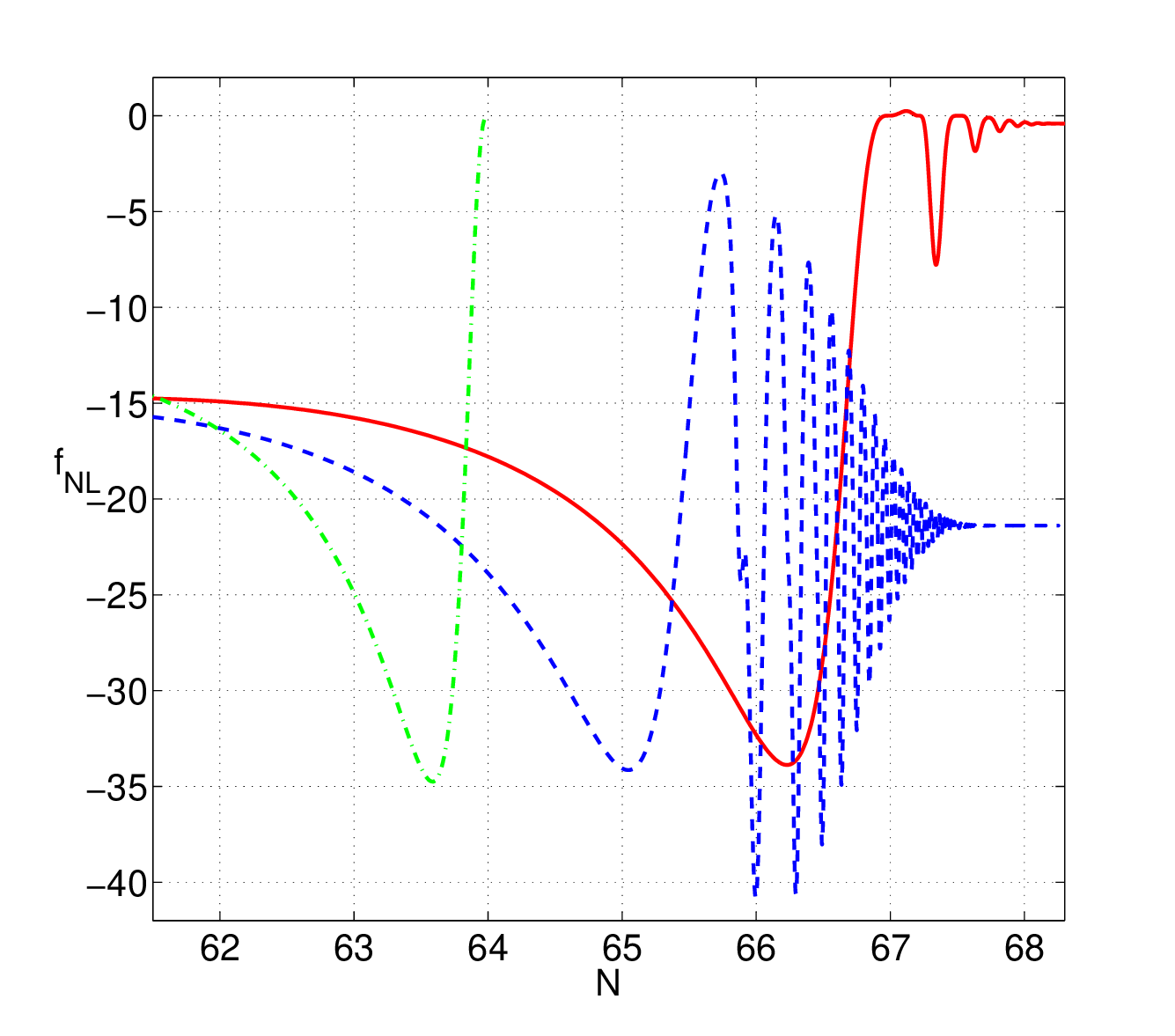}}
	\caption{Upper panel: numerical evolution of $\fNL$
	(solid red line) for the parameter values
    quoted and $\Gamma = (V_0/10)^{1/2}/\Mp$.
    The blue dashed line represents the corresponding plot with
    $\Gamma = (V_0/100)^{1/2}/\Mp$, and the green dot-dashed line
    represents the analytical evolution. 
	The analytic evolution terminates when the $\chi$ field reaches zero,
	because the slow-roll expressions can't evolve
	past this point.
	Lower panel: magnified in the vicinity of the end of inflation.
	The asymptotic
	value of $\fNL$ depends on $\Gamma$,
	and therefore on microphysical details of the reheating phase.}
	\label{figChris1}  
\end{figure}

The initial stage is descent from a ridge, and therefore we expect $\fNL$ to approach negative values. 
This is confirmed in Figs.~\ref{figChris2} and~\ref{figChris1}. As the bundle rolls along the ridge (defined by $\chi = 0$), 
we find that Eq.~\eref{eq:fnl-peak} gives $\fNL \approx -26$, in good agreement with the first negative peak 
in $\fNL$. We have confirmed that the expected scaling with initial conditions is approximately respected.
%-----------------------------------------------------
%-----------------------------------------------
%-------------------------------------------------------------------
With this choice of parameters, a large $\fNL$ is still present as slow-roll breaks down. But because no limiting
trajectory is available, $\fNL$ continues to  evolve---%
and subsequently oscillates wildly, as the fields oscillate about the line $\phi=0$. This does not represent a 
stable attractor: $\phi=0$ is a degenerate vacuum, and the field evolves only along the $\chi$ direction 
during the oscillations. To reach an adiabatic limit we must apply a prescription for reheating. Here, we
adopt a very simple perturbative model in which energy is transferred from the field into a radiation component.
The dynamical equations are
\begin{subequations}
\begin{align}
	\ddot{\phi}_i + 3 H \dot{\phi_i} & = - \Gamma_i \dot{\phi_i}
		- \frac{\partial W}{\partial \phi_i} \\
	\dot{\rho} & = - 4 H \rho +\sum_i \Gamma_i \dot\phi_i^2 ,
\end{align}
\end{subequations}
where $\rho$ is energy density of radiation, and the $\Gamma_i$ represent the decay rate from species $i$. 
We illustrate the effect of reheating in Fig.~\ref{figChris1}. The final value of $\fNL$ is  sensitive to the 
choice of $\Gamma_i$, and hence the time-scale of reheating. We take $\Gamma_i = \Gamma$ for all $i$, 
making reheating begin approximately when $H=\Gamma$ and take place on a uniform density hypersurface. 
A more complicated prescription leads to strong secondary effects which radically alter the value or sign of $\fNL$.
After reheating, if the radiation is the only contribution to the energy density, then the statistics of $\zeta$
at this time will be the ones relevant for observation.
%{\bf For our specific model it appears that the adiabatic value
%of $\fNL$ decays for larger values of $\Gamma$. We do not know if such behaviour is in any way generic, but if it is 
%then it may warrent further investigation.}
Fig.~\ref{figChris1} indicates that these will depend on microphysical
details of the reheating phase, at least through $\Gamma$, but a systematic
understanding is not yet in place.

Our aim has not been to present a realistic model. Rather, we wish to demonstrate that, if no attractor exists within the 
inflationary regime, we must follow the dynamics until all observable quantities stop evolving at the adiabatic limit. 
We can expect the asymptotic value of each observable to be sensitive to this evolution, including the time scale and
details of reheating.

%--------------------------------------------------
\subsection{$\M$-field models}\label{sec:n-field-models}
%--------------------------------------------------

\begin{figure}[htb] 
	\center{\includegraphics[width = 10cm]{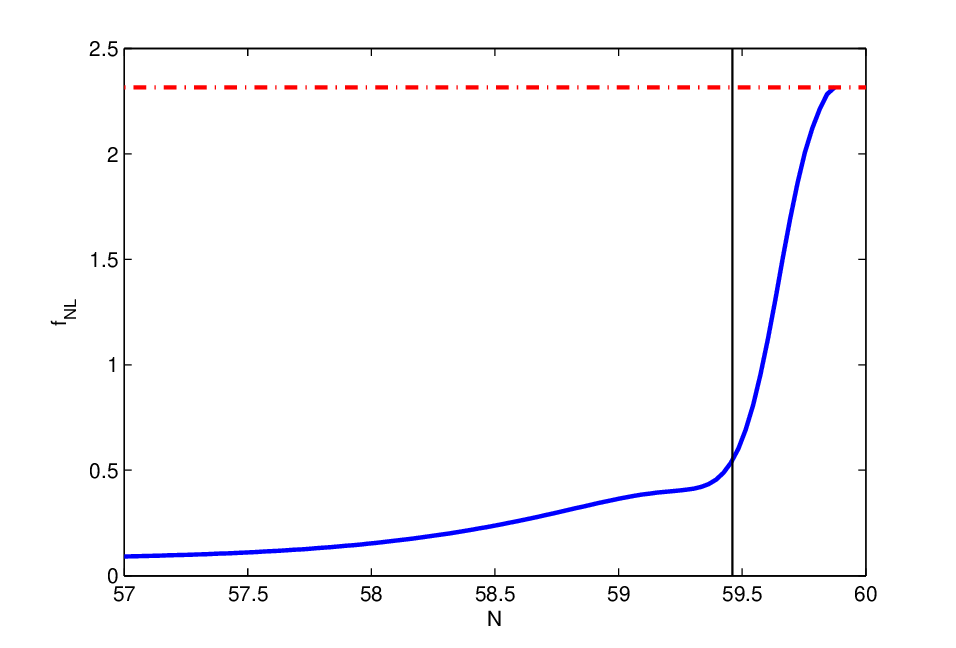}}
    \caption{Evolution of $\fNL$ for the $\M$-axion model, calculated analytically (solid blue line) under the slow-roll
	approximation. The horizontal dot-dashed red line is the asymptotic value computed using the 
	horizon-crossing approximation. (This is unreliable in the present case.) The vertical black line corresponds 
	to $\epsilon = 1$. Since $\fNL$ has not reached the adiabatic limit at this point, this model is an example
	in which the adiabatic limit is reached after slow-roll ends. Therefore, numerical analysis is required to obtain a
	reliable value for $\fNL$.}
	\label{fig:naxion}    
\end{figure}

Similar results naturally apply in models with a larger number of fields. In this section we study the model of Kim {\etal}
\cite{Kim:2010ud} involving many axion fields self-interacting though the potentials 
\begin{equation}
	V_i= \Lambda_i^4 \left(
		1-\cos \frac{2 \pi \phi_i}{f_i}
	\right).
\end{equation}
Where the parameters $\Lambda_i$ and $f_i$ take common values $\Lambda$ and $f$ for each species,
and $f \lesssim \Mp$, this generates naturally large $\fNL$ at the adiabatic limit. Although larger $\fNL$ 
can na\"{\i}vely be obtained by decreasing the $f_i$, it is necessary to simultaneously increase the number 
of fields in order to obtain sufficient inflation. There is another difficulty. As the $f_i$ decrease, the approach 
of $\fNL$ to its asymptotic limit occurs later in the evolution. In Fig.~\ref{fig:naxion} we show one realization 
of this behaviour for $\M = 1800$ and $f_i = \Mp$, with initial conditions for the fields randomly distributed in 
the range $0 < \phi_i < \pi \Mp$. The slow-roll phase ends at latest when $\epsilon = 1$, marked by the
vertical black line. The evolution to the right of this line is not trustworthy and should be replaced by a numerical 
calculation. Unfortunately, owing to the large number of fields we have not been able to perform a non-slow roll 
analysis due to the prohibitive running time of the computation.
%--------------------------------------------------------------------------

There is a specific case where this model can be related exactly to the two-field axion-plus-quadratic 
model of the the previous section: when $\M-1$ fields are initially close to the minimum of the axion potential 
($\phi_i\ll f/2$ for $i=1, \ldots ,\M-1$) with identical initial conditions,  and one field is initially close to its maximum, 
$\phi_{\M} \approx f/2$. In this case the $\M-1$ fields act like a large number of fields with a quadratic potential.
When they all evolve from an identical initial condition, the dynamics of the many fields is completely identical
to the dynamics of a \emph{single} field $\Phi^2=\sum^{\M-1}_{i=1}\phi_i^2$, with a quadratic potential 
of the same mass as the individual $\phi_i$ fields. For $f=\Mp$, this reduces identically to the second of the
two-field axion-plus-quadratic cases studied in~\S\ref{sec:large-natural}. (See Fig.~\ref{figAx2}.)

Finally, it is interesting to note that the $\M$-field axion model---%
for which a large $\fNL$ follows from relatively generic initial conditions---%
is closely related to a two-field model in which generation of large $\fNL$ apparently requires significant
fine-tuning. The tuning appears less dramatic in the original $\M$-field model. It is interesting to conjecture that the 
fine-tuning of initial conditions required to give large $\fNL$ in two-field models may be reduced in models with many fields.

%--------------------------------------------------
\section{Conclusions} \label{sec:conclusions}
%--------------------------------------------------

We have studied the evolution of non-Gaussianity in multiple-field models of inflation.
Unless all isocurvature modes become exhausted before the end of inflation, we find that 
there need not be a unique prediction for $\fNL$. Instead, the final value can depend on
independent details, such as the microphysics of a reheating or preheating phase.
Where the trajectories naturally focus---for example, if inflation ends with all fields settling 
into a minimum of the potential---%
numerical calculations are typically required to determine the precise asymptotic value for $\fNL$.
If there is no natural focusing region then numerical calculations \emph{and} a prescription for 
reheating will be required. This confirms the natural expectation that analytic predictions
\cite{Alabidi:2005qi,Alabidi:2006hg,Vernizzi:2006ve,Battefeld:2006sz} are reliable
only if the flow of power from isocurvature to curvature modes is quenched before the end 
of the slow-roll phase.

If an adiabatic limit is reached without passage through a natural focusing region, perhaps by 
invoking a waterfall transition, then this may occur when the value of $\fNL$ is transiently large.
However, we caution that although our numerical calculations indicate that $\fNL$ can sometimes 
be preserved through a hybrid transition, there does not yet appear to be a precise characterization 
of the conditions required for this to occur. Also, whether the end-point of the waterfall is an adiabatic 
limit may be model dependent. Temporarily ignoring these subtleties, we have shown that descent 
from a ridge or convergence into a valley can result in a significant, dynamical but transient 
enhancement of $\fNL$. The two cases are distinguished by a different sign of the resulting $\fNL$, 
which is inherited from the local $\eta$ parameter.

We have verified that it is possible to construct sum-separable models which exhibit large $\fNL$
even when the adiabatic limit is reached during slow-roll inflation. Therefore there is no correlation 
between large $\fNL$ at the end of inflation and the presence of an inexhausted isocurvature perturbation,
as has occasionally been suggested. On the other hand, we have demonstrated that this is impossible 
for product separable cases, where $\fNL$ is always of order the slow-roll parameters at horizon crossing.

Among the models we have studied is a new two-field model related to the $\M$-field axion model 
\cite{Kim:2010ud}. This exhibits large $\fNL$ at the adiabatic limit when this limit is reached before
the breakdown of slow-roll. In addition, inflation can end gracefully rather than through a sudden transition.
As far as we are aware, this is the first example of such behaviour in the two-field context. This model
explicitly
illustrates that predictions of the $\M$-axion model may be modified if a full numerical calculation for a sufficiently large
number of fields could be performed.

\acknowledgments

We would like to thank Chris Byrnes, Andrew Liddle and Karim Malik for helpful discussions.
JE is supported by a Science and Technology Facilities Council Studentship.
DJM is supported by the Science and Technology Facilities Council grant ST/H002855/1. 
DS was supported by the Science and Technology Facilities Council [grant numbers ST/F002858/1 and ST/I000976/1].

\appendix
\section{Detailed calculations for separable potentials} \label{sec:appendix}

\subsection{Calculating $N_{,i}$}

Under the assumption of slow-roll and monotonicity ($\dot \phi_k < 0$), the number of e-folds can be written with
the field $\phi_k$ as a time variable. Taking the functional derivative of this integral generates three
components---two ``boundary terms'' evaluated on the initial (`$*$') and final (`$c$') slices, and a ``path term'',
\be
\label{eq:initial_expansion}
N_{,i} \Mp^2 = \left. \frac{W}{W_{,k}} \right|_* \delta_{ik} - 
\left. \frac{W}{W_{,k}} \right|_c \frac{\partial \phi_k^c}{\partial \phi_i^*}-
\int_*^c \frac{\partial}{\partial \phi_i^*} \left( \frac{W}{W_{,k}} \right) \d \phi_k  .
\ee
The summation convention is not used. Physical quantities are independent of $k$, which may be chosen
arbitrarily. Employing the notation $S = \sum_{i=1}^{\M} V_i$ and 
$P = \prod_{i=1}^{\M} V_i$, where $V_1 = V_1 (\phi_1)$ are \emph{dimensionless}, we restrict attention to potentials of the 
form $W = \Mp^4 \, F(S)$ and $W = \Mp^4 \, G(P)$, where $F$ and $G$ are arbitrary functions. We refer to these as
sum- and product-separable potentials respectively.

For the product-separable potential $W= \Mp^4 \, G(P)$ we evaluate the path term and the final boundary term 
in Eq. \eqref{eq:initial_expansion}. Using the slow-roll parameters $\sqrt{ 2\ep_i} = \Mp \, G'P V_i' /G V_i$
and also defining $u_i = \ep_i / \ep$ we find
\be
N_{,i} = \frac{1}{\Mp ^2} \left. \frac{V_i}{V_i'}\right|_* \left( \left. \frac{G u_i}{G' P} \right|_c 
- \int_*^c \frac{\partial}{\partial \phi_i} \left( \frac{G}{G'P} \right) \d \phi_i \right).
\ee
This applies for arbitrary $G$. Note that any dependence on $k$ has disappeared. If the ratio $G/G'P$ 
depends on $P$, then the integral requires knowledge of the variation of $V_j$ with $\phi_i$.
There are at least two cases in which the $P$-dependence is lost: if the integrand is either zero or constant.
If the integrand is zero then we have $G/G'P = A$ for some constant $A$. This gives the general solution
$G = AP^B$, where the constants $A$ and $B$ can be absorbed into a redefinition of the potential. This yields
\be
\label{eq:ps_Ni_1}
N_{,i} = \frac{1}{\Mp} \frac{u_i^c}{\sqrt{2 \ep_i^*}}, \quad W = \Mp^4 \, P.
\ee
If the integrand is constant this implies $G = (B \ln P+D)^{1/A}$, where $A,B,D$ are constants of which
$B$ and $D$ can be eliminated by a further redefinition. We conclude that a general potential yielding a 
$P$-independent integrand can be expressed as $W = \Mp^4 \, (\ln P)^{1/A}$ and gives 
\be
\label{eq:ps_Ni_2}
N_{,i} = \frac{A}{\Mp^2} \left. \frac{V_i}{V_i'}\right|_* \left( \ln V_i^* - \ln V_i^c + (\ln P)^c u_i^c \right),
\quad W = \Mp^4 \, (\ln P)^{1/A}.
\ee
An identical procedure applies to the sum-separable potential, leading to
\bea
\label{eq:ss_Ni_1}
N_{,i} &=& \frac{1}{\Mp} \frac{u_i^c}{\sqrt{2 \ep_i^*}},
\quad W = \Mp^4 \, e^S \\
\label{eq:ss_Ni_2}
N_{,i} &=& \frac{A}{\Mp^2 \, {V_i'}^*} \left( V_i^* - V_i^c + S^c u_i^c \right),
\quad
W = \Mp^4 \, S^{1/A}.
\eea

\subsection{Correspondence between different separable potentials}

There is a correspondence between Eqs.~\eqref{eq:ss_Ni_1} and~\eqref{eq:ps_Ni_1}, and between 
Eqs.~\eqref{eq:ss_Ni_2} and~\eqref{eq:ps_Ni_2} which was first pointed out by Wang \cite{Wang:2010si} for two field potentials.
Redefining $\ln V_i \to V_i$ turns a $W=\Mp^4 \, P$ potential into a $W=\Mp^4 \, e^S$ potential, and redefining $e^{V_i} \to V_i$ transforms 
the potential $W= \Mp^4 \, S^{1/A}$ into the form $W=\Mp^4 \, (\ln P)^{1/A}$. Therefore, it is unnecessary to proceed with all four classes of potential.
As an independent pair of potentials, we choose $W=\Mp^4 \, P$ and $W= \Mp^4 \, S^{1/A}$.

For $W=\Mp^4 \, P$:
\be
\label{eq:ps_Ni}
\frac{\partial \phi_k^c}{\partial \phi_i^*} = \sqrt{\frac{\ep_k^c}{\ep_i^*}} \left( \delta_{ik} - u_i^c \right) ~~~{\rm and}~~~ 
N_{,i} = \frac{1}{\Mp} \frac{u_i^c}{\sqrt{2 \ep_i^*}}.~~~~~~~~~~~~~~~
\ee
For $W= \Mp^4 \, S^{1/A}$:
\be
\label{eq:ss_Ni_change}
~~~~~~~~~~~~~~~\frac{\partial \phi_k^c}{\partial \phi_i^*} = \frac{S^c}{S^*} \sqrt{\frac{\ep_k^c}{\ep_i^*}} \left( \delta_{ik} - u_i^c \right) ~~~{\rm and}~~~ 
N_{,i} = \frac{1}{\Mp \, S^* \sqrt{2 \ep_i^*}} \left( V_i^* - V_i^c + S^c u_i^c \right).
\ee
These formulae have previously appeared in Refs.~\cite{Battefeld:2006sz,Byrnes:2008wi,Wang:2010si}.

\subsection{$N_{,ij}$ and $ \fNL$}

To calculate higher-order statistics we require the second derivatives $N_{,ij}$. 
It proves convenient to introduce new dimensionless slow-roll parameters
\bea
a_i &=& \Mp \, \frac{\partial}{\partial \phi_i} \ln \left( \frac{W}{\Mp^4} \right) = \Mp \, \frac{W_{,i}}{W} = \sqrt{2 \ep_i} \\
b_{ij} &=& \Mp^2 \, \frac{\partial^2}{\partial \phi_i \partial \phi_j} \ln \left( \frac{W}{\Mp^4} \right) = 
\Mp^2 \, \left( \frac{W_{,ij}}{W} - \frac{W_{,i}W_{,j}}{W^2} \right) = \eta_{ij} - 2\sqrt{\ep_i \ep_j} .
\eea
These are elements of a vector $\bm a$ and matrix $\bm b$ respectively. 
Note that $\bm b$ is diagonal for the product-separable potential $W=\Mp^4 \, P$ 
and so we rewrite $N_{,i} = u_i^c / \Mp \, a_i^* $ with $u_i = a_i^2 / |{\bm a}|^2$.
A simplification occurs if the limiting  trajectory is a straight line in field space,
which implies
\be
\left. \fNL \right|_{\rm straight}= - \frac{5}{6} \frac{1}{|{\bm q}^c|^4} \sum_{i=1}^{\M} (q_i^c)^3 \frac{b_{ii}^*}{a_i^*}
\ee
where the vector $\bm q^c$ has elements $q_i^c = N_{,i}$.
If the limiting trajectory lies along the $\phi$ axis, then $\fNL = - \frac{5}{6} b_{\phi \phi}^*$.
In this case, the sign of $\fNL$ is given by the mass of the field at horizon crossing but the 
magnitude is slow-roll suppressed. 
If we do not make the simplifying assumption that the limiting trajectory is a straight line, then
\bea
\Mp^2 \, N_{,ij} &=& \frac{q_i^c}{a_i^*} \left( 2 b_{ii}^c - b_{ii}^* \right) \delta_{ij} - 
2 q_i^c q_j^c \left. \left( b_{ii} + b_{jj}  - \frac{\bm{a \cdot b \cdot a}}{|\bm a|^2} \right) \right|_c \\
\frac{6}{5} \fNL &=& \frac{1}{|{\bm q^c}|^4} \sum_{i=1}^{\M} \left( \frac{{q_i^c}^3}{a_i^*} (2 b_{ii}^c - b_{ii}^*)\right)
- 4 \left. \frac{\bm{q \cdot b \cdot q}}{|{\bm q}|^2} \right|_c + 2 \left. \frac{\bm{a \cdot b \cdot a}}{|{\bm a}|^2}\right|_c.
\eea

For the potential $W = \Mp^4 \, S^{1/A}$ we have $N_{,i} = \left( V_i^* - V_i^c + S^c u_i^c\right) / \Mp \, a_i^* S^*$ . 
In the simpler situation where $u_i^c$ is a constant, we find
\bea
\label{eq:ss_Nij_straight}
\Mp^2 \, \left. N_{,ij} \right|_{\rm straight} &=& A \delta_{ij} - q_i^c \left( A a_j^* + \frac{b_{ij}^*}{a_i^*} \right)
- A \frac{S_c^2}{S_*^2} \frac{{a_i^c}^2}{a_i^* a_j^*} (\delta_{ij} - u_j^c) \\
\label{eq:ss_fnl_straight}
\left. \fNL \right|_{\rm straight} &=& \frac{5}{6}\frac{A}{|{\bm q}^c|^2}
-  \frac{5}{6}\frac{1}{|{\bm q}^c|^4} \sum_{i,j=1}^{\M} \! \! \left(
{q_i^c}^2 q_j^c \left( A a_i^* + \frac{b_{ij}^*}{a_i^*} \right) + A \frac{S_c^2}{S_*^2} \frac{{a_i^c}^2 q_i^c q_j^c}{a_i^* a_j^*}  (\delta_{ij} - u_j^c) \right) ~~~~~~~~
\eea

Dropping the assumption that $u_i^c$ is constant leads to additional terms of the form
\bea
\label{eqA:ss_Nij}
N_{,ij}&=& \left. N_{,ij} \right|_{\rm straight} 
+\frac{2}{\Mp^2} \frac{S_c^2}{S_*^2} \frac{M_{ij}^c}{a_i^* a_j^* |{\bm a}^c|^2} \\
\fNL &=& \left. \fNL \right|_{\rm straight} 
+ \frac{5}{3} \frac{S_c^2}{S_*^2} \frac{1}{|{\bm q}^c|^4 |{\bm a}^c|^2} 
\sum_{i,j=1}^{\M} \frac{M_{ij}^c q_i^c q_j^c}{a_i^* a_j^*} \\
M_{ij} &=& a_i b_{ij} a_j - a_i u_j (\bm{ b \cdot a})_i - a_j u_i (\bm{ b \cdot a})_j + u_i u_j (\bm{ a \cdot b \cdot a})  .
\label{eqA:ss_fnl}
\eea

\bibliography{adiabaticity}{}
\bibliographystyle{JHEPmodplain}

\end{document}